\documentclass[aps,pre,twocolumn,showpacs,superscriptaddress]{revtex4-1}
\pdfoutput=1
\usepackage{graphicx}% Include figure files
\usepackage{dcolumn}% Align table columns on decimal point
\usepackage{bm}% bold math
\usepackage{amsmath,amssymb,mathtools}
\usepackage{color}
\definecolor{darkblue}{rgb}{0,0,0.6}
\definecolor{darkred}{rgb}{0.6,0,0}
\definecolor{darkgreen}{rgb}{0,0.6,0}

\usepackage[colorlinks=true,urlcolor=darkblue,citecolor=darkblue,linkcolor=darkred,hyperfootnotes=false]{hyperref}

\usepackage{setspace}

\makeatother

\begin{document}

\title{Generic long-range interactions between passive bodies in an active fluid}

\author{Yongjoo Baek}
\email{y.baek@damtp.cam.ac.uk}
\affiliation{Department of Physics, Technion -- Israel Institute of Technology, Haifa 32000, Israel}
\affiliation{DAMTP, Centre for Mathematical Sciences, University of Cambridge, Cambridge CB3 0WA, United Kingdom}

\author{Alexandre P. Solon}
\affiliation{Department of Physics, Massachusetts Institute of Technology, Cambridge, Massachusetts 02139, USA}

\author{Xinpeng Xu}
\affiliation{Department of Physics, Technion -- Israel Institute of Technology, Haifa 32000, Israel}
\affiliation{Department of Physics, Guangdong Technion -- Israel Institute of Technology, Shantou, Guangdong 515063, P. R. China}

\author{Nikolai Nikola}
\affiliation{Department of Physics, Technion -- Israel Institute of Technology, Haifa 32000, Israel}

\author{Yariv Kafri}
\affiliation{Department of Physics, Technion -- Israel Institute of Technology, Haifa 32000, Israel}

\date{\today}
 
\begin{abstract}
Because active particles break time-reversal symmetry, a single non-spherical body placed in an active fluid generates currents. We show that when two or more passive bodies are placed in an active fluid these currents lead to long-range interactions. Using a multipole expansion we characterize their leading-order behaviors in terms of single-body properties and show that they decay as a power law with the distance between the bodies, are anisotropic, and do not obey an action--reaction principle. The interactions lead to rich dynamics of the bodies, illustrated by the spontaneous synchronized rotation of pinned non-chiral bodies and the formation of traveling bound pairs. The occurrence of these phenomena depends on tunable properties of the bodies, thus opening new possibilities for self-assembly mediated by active fluids.
\end{abstract}


\maketitle

Active matter is a class of nonequilibrium systems in which energy is converted into systematic motion on a microscopic scale~\cite{RamaswamyARCMP2010}. They have attracted much attention~\cite{MarchettiRMP2013,BechingerRMP2016} due to a host of interesting physical phenomena~\cite{VicsekPRL1995,VicsekPhysRep2012,TonerAnnPhys2005,CatesARCMP2015,HenkesPRE2011,SheinmanPRL2015}, their relevance to many biological systems~\cite{ChowdhuryPhysLifeRev2005,KolomeiskyARPC2007,JulicherPhysRep2007,ChowdhuryPhysRep2013}, and their potential use for self-assembly applications~\cite{SotoPRL2014,*SotoPRE2015}. They have also been suggested as tools for novel engineering applications -- for example, active fluids have been used to power microscopic gears~\cite{AngelaniNJP2010,KaiserPRL2014,KaiserEPJST2015,AngelaniPRL2009,SokolovPNAS2010,DiLeonardoPNAS2010,ZhangEPL2013}. This results from the fact that, when an asymmetric body is immersed in a fluid with broken time-reversal symmetry, it experiences a net force~\cite{ReichhardtARCMP2017,MalloryPRE2014b,YanJFM2015} which is coupled to the generation of ratchet-like currents~\cite{TailleurEPL2009,NikolaPRL2016}. 

In this Letter we study passive bodies immersed in an active fluid. We show that the ratchet-like currents generated by each body give rise to forces and torques which decay as a power law with distance, are anisotropic, and do not
obey an action--reaction principle. Using a multipole expansion, the leading-order behavior of the 
interactions can be expressed in terms of single-body
quantities that can be measured independently in experiments or numerical simulations. Moreover, by designing the two bodies one can control the amplitude and polarity of the interactions between them. This leads to a host of interesting dynamical phenomena of which we illustrate two: the spontaneous synchronized rotations of pinned rotors and the formation of traveling bound pairs. Our results suggest a new method for self-assembly by embedding passive bodies in an active fluid.

We stress that the interactions studied here exist even between
non-moving bodies and are therefore distinct from usual hydrodynamic
interactions~\cite{HappelBrenner1983}. They are also different from
thermal Casimir interactions~\cite{FisherCRAS1978,KardarRMP1999}, because they do not rely on correlations between the fluid particles and are present even in a dilute active fluid.

{\em Model.} --- We base our study on a common model of an active fluid consisting of $N$ point-like particles, which do not interact among themselves and self-propel at a constant speed $v$ in two dimensions.
%We base our study on a common model of an active fluid consisting of $N$ point-like noninteracting particles, self-propelled at a constant speed $v$ in two dimensions.
The position $\mathbf{r}_i$ and the orientation $\theta_i$ of active particle $i$ evolve according to the overdamped Langevin equations
\begin{align} \label{eq:langevin}
\dot{\mathbf{r}}_i &= v
\mathbf{e}_{\theta_i} - \mu \sum_j \bm\nabla V_j + \sqrt{2D_t}\bm
\eta_i\;, \nonumber\\
\dot{\theta}_i &= \sqrt{2D_r}\xi_i\;.
\end{align}
Here $\mathbf{e}_{\theta_i} \equiv (\cos\theta_i,\sin\theta_i)$ is
the heading of particle $i$, $\mu$ is its mobility, $D_t$ and $D_r$ are translational and
rotational diffusivities, and $\bm \eta_i$ and $\xi_i$ are Gaussian
white noises of unit variance. The presence of body $j$ in the active fluid is described by a potential $V_j$ describing the interaction between each active particle and the
passive body~$j$. The dots denote derivatives with
respect to time. In addition, we allow the particles to randomize
their orientation at a constant tumbling rate $\alpha$. This dynamics
encompass the two well-studied models of run-and-tumble
particles (RTPs, with $\alpha \neq 0$ and
$D_r = 0$)~\cite{SchnitzerPRE1993} and active Brownian particles
(ABPs, having $\alpha = 0$ and
$D_r \neq 0$)~\cite{SchweitzerPRL1998,RomanczukEPJST2012}. There has
been much recent
progress~\cite{MalloryPRE2014a,YangSM2014,TakatoriPRL2014,FilySM2014,SolonPRL2015,GinotPRX2015,TakatoriPRE2015,YanJFM2015,YanSM2015,SolonNP2015,WinklerSM2015,SpeckPRE2016,NikolaPRL2016,FilyArXiv2017,SandfordArXiv2017,SandfordPRE2017}
in the characterization of forces in this class of systems and we
build upon it. Note that the model falls into the class of dry active systems, which do
not conserve momentum. As such it is best suited for describing the
dynamics of particles next to a surface, for example those of vibrated
granular monolayers~\cite{DeseignePRL2010,DeseigneSM2012,JunotPRL2017}
or gliding bacteria~\cite{PeruaniPRL2012}.

\begin{figure*}
\includegraphics[width=0.99\textwidth]{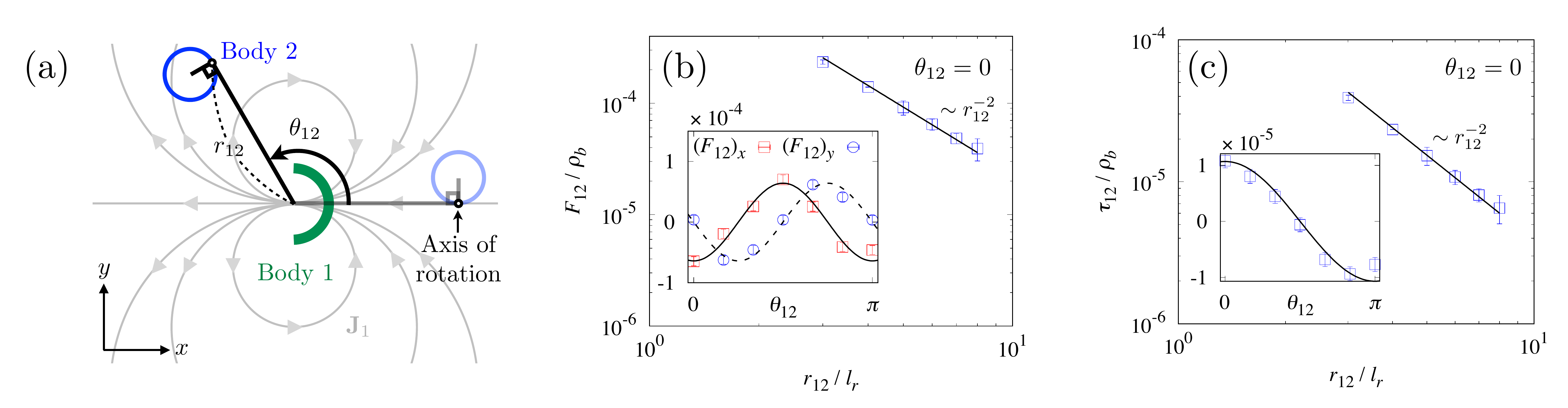}
\caption{\label{fig:fig1} Dipole contributions to far-field
interactions. (a) The numerical setup used to verify the theoretical predictions. A circular body is placed at different distances $r_{12}$ and angles $\theta_{12}$ away from a semicircular body which generates a dipole current. (b) The force normalized by the bulk density as a function of $r_{12}/l_r$, with $l_r$ the run length, for $\theta_{12}=0$. (Inset) The dependence of the force components $(F_{12})_{x,y}$ in the $x$- and $y$-directions as a function of $\theta_{12}$. (c) The torque applied by the
semicircle on a circle which is held tangent to ${\bf r}_{12}$ and pinned at the contact point as $\theta_{12}$ is varied, see (a). The lines in (b) and (c) correspond to the theory with no fitting parameters. All the parameters and units of the simulations are specified in Appendix~\ref{app:parameters}.}
\end{figure*}

{\em A single passive body.} --- We first discuss the effects of a single passive body on an active fluid. For future reference, we maintain the index~$j$, although in this case body~$j$ is the only passive body in the system. Using standard methods~\cite{CatesEPL2013,SolonEPJST2015} (also detailed in Appendix~\ref{app:density_current}), Eq.~\eqref{eq:langevin} leads to an exact equation for the active particle density $\rho_j$ ($j$ marking the dependence on $V_j$) in the steady state
\begin{align} \label{eq:poisson} D_\text{eff}\nabla^2\rho_j &=
-\mu\bm\nabla\cdot(\rho_j\bm\nabla V_j) + \sum_{a,\,b}\partial_a\partial_b (\mathbb{G}_j)_{ab}\;.
\end{align}
Here $D_\text{eff}\equiv D_t + v\,l_r/2$ is the effective diffusion
constant of the active particles, $l_r \equiv v/(\alpha+D_r)$ their
run length, indices $a$ and $b$ run over the Cartesian coordinates $\{x,\,y\}$, and $\mathbb{G}_j$ is a rank~$2$ tensor containing information about active particle orientations. In the far-field limit (at distances $r$ much larger than the diameter of the body and the run length $l_r$), the solution of Eq.~\eqref{eq:poisson} can be written to dipole order in a multipole expansion as
\begin{align} \label{eq:dipole_density}
\rho_j(\mathbf{r}) =
\rho_b +
\frac{\beta_\text{eff}}{2\pi}\,\frac{\mathbf{r}\cdot\mathbf{p}_j}{r^2}
+ O(r^{-2})
\;,
\end{align} 
where $\rho_b$ denotes the bulk density of active particles and
$\beta_\text{eff} \equiv \mu/D_\text{eff}$ their effective inverse
temperature. The dipole moment $\mathbf{p}_j$, obtained as
\begin{align} \label{eq:dipole_moment}
\mathbf{p}_j = -\int
d^2\mathbf{r}'\,\rho_j\bm\nabla'V_j\;,
\end{align}
is equal to the total force exerted by body~$j$ on the active
particles. Alternatively, $-\mathbf{p}_j$ is the propulsion force
applied by the active particles on body~$j$.
%As previously noted, it is nonzero for asymmetric bodies due to ratchet-like effects.
We stress that at this order Eqs.~\eqref{eq:dipole_density} and \eqref{eq:dipole_moment} are exact, with $\mathbb{G}_j$ in Eq.~\eqref{eq:poisson} only contributing to higher-order multipoles in the far field. While $\mathbf{p}_j$ is easily measurable from Eq.~\eqref{eq:dipole_moment}, its first-principle calculation is difficult due to complex near-field effects of $\mathbb{G}_j$. However, as shown in Appendix~\ref{app:perturbation}, it can be perturbatively obtained for shallow potentials $\beta _\text{eff} |\nabla V_j|\ll 1$, explicitly confirming that asymmetric $V_j$ induces $\mathbf{p}_j \neq \mathbf{0}$.

The associated far-field current density is dominated by the diffusive component
\begin{align} \label{eq:dipole_current}
\mathbf{J}_j(\mathbf{r})
  \simeq -D_\text{eff}\bm\nabla\rho_j \simeq
  -\frac{\mu}{2\pi}\left[\frac{\mathbf{p}_j}{r^2} -
    \frac{2(\mathbf{r}\cdot\mathbf{p}_j)\mathbf{r}}{r^4}\right]\;,
\end{align}
which resembles the electric field of a charge dipole. The forms of $\rho_j$ and $\mathbf{J}_j$ are equivalent to those of the density and current fields generated by a local pump applying a point force $\mathbf{p}_j$ on a passive diffusive medium~\cite{SadhuPRE2011}. In this sense an asymmetric passive body in an active fluid acts like a pump, although its power is supplied not by an external source, but by the active particles themselves.

{\em Forces between passive bodies.} --- We now consider two passive
bodies described by potentials $V_1$ and $V_2$, with dipole moments
(in isolation) ${\mathbf p}_1$ and ${\mathbf p}_2$, and position vectors 
$\mathbf{R}_1$ and $\mathbf{R}_2$. We set
$\mathbf{R}_2 = \mathbf{0}$ and $\mathbf{R}_1 = \mathbf{r}_{12}$ and work in the far-field limit where $r_{12}=|\mathbf{r}_{12}|$ is much larger than the run length $l_r$ and the diameters of the bodies. Denoting by $\rho$ the steady-state density field of active particles, the force applied by the active particles on body~$j$ is given by $\int d^2\mathbf{r} \,\rho\bm\nabla V_j$, with $j \in \{1,\,2\}$. We define the force applied by body~$1$ on body~$2$ as $\mathbf{F}_{12} = \mathbf{p}_2 + \int d^2\mathbf{r} \,\rho\bm\nabla V_2$, which is the change in the force acting on body~$2$ due to the presence of body~$1$ (recall that $-\mathbf{p}_2$, given by Eq.~\eqref{eq:dipole_moment}, is the force acting on isolated body~$2$). This stems from the change of $\rho$ near body~$2$ induced by body~$1$, which can be expressed as a series expansion in $r_{12}^{-1}$ (see Appendix~\ref{app:interactions} for a detailed derivation). It is convenient to separate the total force into two components $\mathbf{F}_{12} = \mathbf{F}^a_{12} + \mathbf{F}^s_{12}$, where $\mathbf{F}^a_{12}$ acts only on asymmetric bodies with nonzero $\mathbf{p}_j$, while $\mathbf{F}^s_{12}$ is present even for symmetric bodies with $\mathbf{p}_j = 0$. Then we find
\begin{align}
\mathbf{F}^a_{12} &= -\frac{\beta_\text{eff}}{2\pi\rho_b}\frac{\mathbf{r}_{12}\cdot\mathbf{p}_1}{r_{12}^2}\,\mathbf{p}_2 + O(r_{12}^{-2}) \;,\label{eq:F12_nsym} 
\\ \mathbf{F}^s_{12} &= \frac{\mathbb{R}_2\mathbf{J}_1(\mathbf{r}_{12})}{\rho_b} +O(r_{12}^{-3}) \;. \label{eq:F12_sym}
\end{align}
Here $\mathbb{R}_2$ is the inverse mobility tensor of body~$2$ due to the
active particles. It is measured by placing
body~$2$ alone in an active fluid of average density $\rho_b$, through which a boundary-driven diffusive current $\rho_b\mathbf u$ is flowing. Then $\mathbb{R}_2$ is calculated from
\begin{align} \label{eq:R2} (\mathbb{R}_2)_{ab}
=\left. \frac{\partial}{\partial u_b}\left[ {\mathbf F}_2^{(\mathbf{u})} \cdot \mathbf{e}_a \right]\right|_{\mathbf{u}=\mathbf{0}}\;,
\end{align}
where indices $a$ and $b$ stand for Cartesian coordinates $\{x,y\}$, $\mathbf{e}_a$ is a unit vector in the $a$-direction, and $\mathbf{F}_2^{(\mathbf{u})}$ is the steady-state force on the body.
Finally, $\mathbf{J}_1$ in
Eq.~\eqref{eq:F12_sym} is given by Eq.~\eqref{eq:dipole_current}. At this order, the interactions between the two bodies are thus completely determined by
$\mathbf{p}_1$, $\mathbf{p}_2$, and $\mathbb{R}_2$, which are single-body quantities
that can be measured independently.

We can understand Eqs.~\eqref{eq:F12_nsym} and \eqref{eq:F12_sym} intuitively in terms of the density and current fields produced by body~$1$ alone. First, we note that body~$2$ experiences a propulsion force $-\mathbf{p}_2$ in the absence of body~$1$. Due to the mutual independence of active particles, the propulsion force is proportional to the bulk density $\rho_b$. With body~$1$ added its dipole density field changes the effective bulk density felt by body~$2$ from $\rho_b$ to $\rho_1(\mathbf{r}_{12})$, given in Eq.~\eqref{eq:dipole_density}. This leads to Eq.~\eqref{eq:F12_nsym} with $\mathbf{F}^a_{12} \sim r_{12}^{-1}$. Therefore $\mathbf{F}^a_{12}$ only induces a correction to the speed of the body.

Meanwhile, $\mathbf{F}^s_{12} \sim r_{12}^{-2}$ stems from the force on body $2$ due to the current field $\mathbf{J}_1(\mathbf{r}_{12}) \sim r_{12}^{-2}$ generated by body~$1$ in accordance with Eq.~\eqref{eq:dipole_current}. At large $r_{12}$ the induced force can be linearized as $\mathbb{R}_2\mathbf{J}_1(\mathbf{r}_{12})/\rho_b$. Note that $\mathbf{F}_{12}^s$ can change the propulsion direction of body~$2$. In Fig.~\ref{fig:fig1} we present a measurement of the force $\mathbf{F}_{12}$ on a circular body (so that ${\bf p}_2=\mathbf{0}$). In this case $\mathbb{R}_2$ is proportional to the identity matrix and we evaluated it numerically. The results agree nicely with the theory using no fitting parameters, although on a reduced range because of numerical limitations.

{\em Torques between passive bodies.} --- The torque $\bm\tau_{12}$
exerted by body~$1$ on body~$2$ can be obtained using the same
approach. We denote by
$\bm \tau_j=\int d^2 \mathbf{r}' \,
\rho_{j}(\mathbf{r}')\left(\mathbf{r}'-\mathbf{R}_j
\right)\times\bm\nabla' V_j$
the self-torque on an isolated body~$j$ with respect to the reference position $\mathbf{R}_j$. It is useful to decompose the result into a correction to the self-torque $\bm \tau_{12}^a$, which is dominant when $\bm\tau_2 \neq \mathbf{0}$, and a sub-leading contribution $\bm \tau_{12}^s$, which induces a torque even when $\bm\tau_2 = \mathbf{0}$. We find
\begin{align}
	 \bm \tau_{12}^a &=
\frac{\beta_\text{eff}}{2\pi\rho_b}\frac{\mathbf{r}_{12}\cdot\mathbf{p}_1}{r_{12}^2}\,\bm
\tau_{2} +O(r_{12}^{-2})\;, \label{eq:tau12_nsym} \\
\bm \tau_{12}^s &= \frac{\bm\gamma_2}{\rho_b}\times\mathbf{J}_1(\mathbf{r}_{12}) +
O(r_{12}^{-3})\;, \label{eq:tau12_sym}
\end{align} 
where the vector $\bm\gamma_2$, similarly to $\mathbb{R}_2$, characterizes the response of isolated body~$2$ to a diffusive current $\rho_b \mathbf{u}$ carried by an active fluid of mean density $\rho_b$. The vector is calculated from the steady-state torque, ${\bm \tau}_2^{({\mathbf u})}$, exerted on body~$2$ according to
\begin{align} \bm\gamma_2 = \left. \left[ \bm\nabla_\mathbf{u} \times
 {\bm \tau}_2^{({\mathbf u})}
\right] \right|_{\mathbf{u}=\mathbf{0}} \;. \label{eq:gamma}
\end{align}
As with Eqs.~\eqref{eq:F12_nsym} and \eqref{eq:F12_sym}, $\bm \tau_{12}^a$ results from a local shift in the density, and $\bm \tau_{12}^s$ from the current. The latter tends to align $\bm\gamma_2$ with the current. In Fig.~\ref{fig:fig1}, our predictions are compared with simulations which measure the torque exerted by a semicircle (body~$1$) on a circle (body~$2$) held at its edge, with $\bm\gamma_2$ evaluated numerically.

A few comments on the properties of the interactions are in order. First, even in the presence of three or more passive bodies, at large mutual distances the interactions are still dominated by pairwise components. Second, going one order higher in the multipole expansion, one finds that two rod-like bodies interact through quadrupole moments. A previous study~\cite{RayPRE2014} on the same setup considered the interaction a near-field effect, but we predict this to be a long-range force decaying as $r_{12}^{-3}$. A numerical support is provided by Fig.~\ref{fig:figS2}. Finally, an extension of the analysis to dimensions $d > 2$ yields ${\bf F}_{12}^a \sim r_{12}^{-(d-1)}$ and  ${\bf F}_{12}^s \sim r_{12}^{-d}$, with corresponding changes to the torques.

We note that the interactions discussed above are anisotropic and do not satisfy the action--reaction principle. For passive bodies allowed to move in the active fluid, these features lead to a host of interesting dynamical phenomena. Assuming overdamped bodies, we take
\begin{align} \label{eq:body_motion}
\dot{\mathbf{R}}_j &= \mu_j^T \int d^2\mathbf{r}\, \rho(\mathbf{r},t)\,\bm\nabla V_j \;, \nonumber\\
\dot{\Theta}_j &= \mu_j^R\int d^2\mathbf{r}\, \rho(\mathbf{r},t)\,\left[(\mathbf{r}-\mathbf{R}_j)\times\bm\nabla V_j\right]\cdot \mathbf{e}_z \;,
\end{align}
where $\mu_j^T$ and $\mu_j^R$ are the translational and rotational
mobilities of body~$j$, $\Theta_j$ is an angle giving its orientation
with respect to a fixed axis of reference, and $\mathbf{R}_j$ is a position vector. For simplicity we consider bodies for which the position vector $\mathbf{R}_j$ can be chosen so that no off-diagonal mobilities couple the translational and rotational degrees of freedom. The extension to other cases is straightforward. The results derived above are applicable when the mobilities are small enough so that an adiabatic limit holds: at
each instant, the system can be considered to be in a steady state
with fixed body positions. We do not specify direct (short-range)
interactions between the bodies since here we consider only
far-field effects. Interestingly, we show that properties of the bodies can be tuned to
lead to distinct phenomena.

For concreteness, we consider pairs of
{\it semicircles}. A semicircle traps active particles approaching
from the concave side (`rear'), while those on the convex side
(`front') easily slide past the body. This creates a backward current
of active particles, associated with an opposing
force~\cite{NikolaPRL2016} which propels the semicircle forward (see Fig.~\ref{fig:fig1}a). In
the terminology introduced above, a semicircle has a nonzero dipole
moment $\mathbf{p}_j$ pointing in the backward direction. We focus on two types of semicircles that
rotate around $\mathbf{R}_j$ either at the apex (type~$A$) or at the center of the circle
(type~$C$). In experiments, this could be achieved by properly designing the
bodies. Most importantly for their dynamics, $\bm \gamma_j$ is parallel to $\mathbf{p}_j$ for type~$C$ bodies and antiparallel
for type~$A$ bodies. In both
cases, $\mathbf{R}_j$ is on the symmetry axis so that $\bm \tau_A=\bm \tau_C=\mathbf{0}$.

\begin{figure}
\includegraphics[width=0.95\columnwidth]{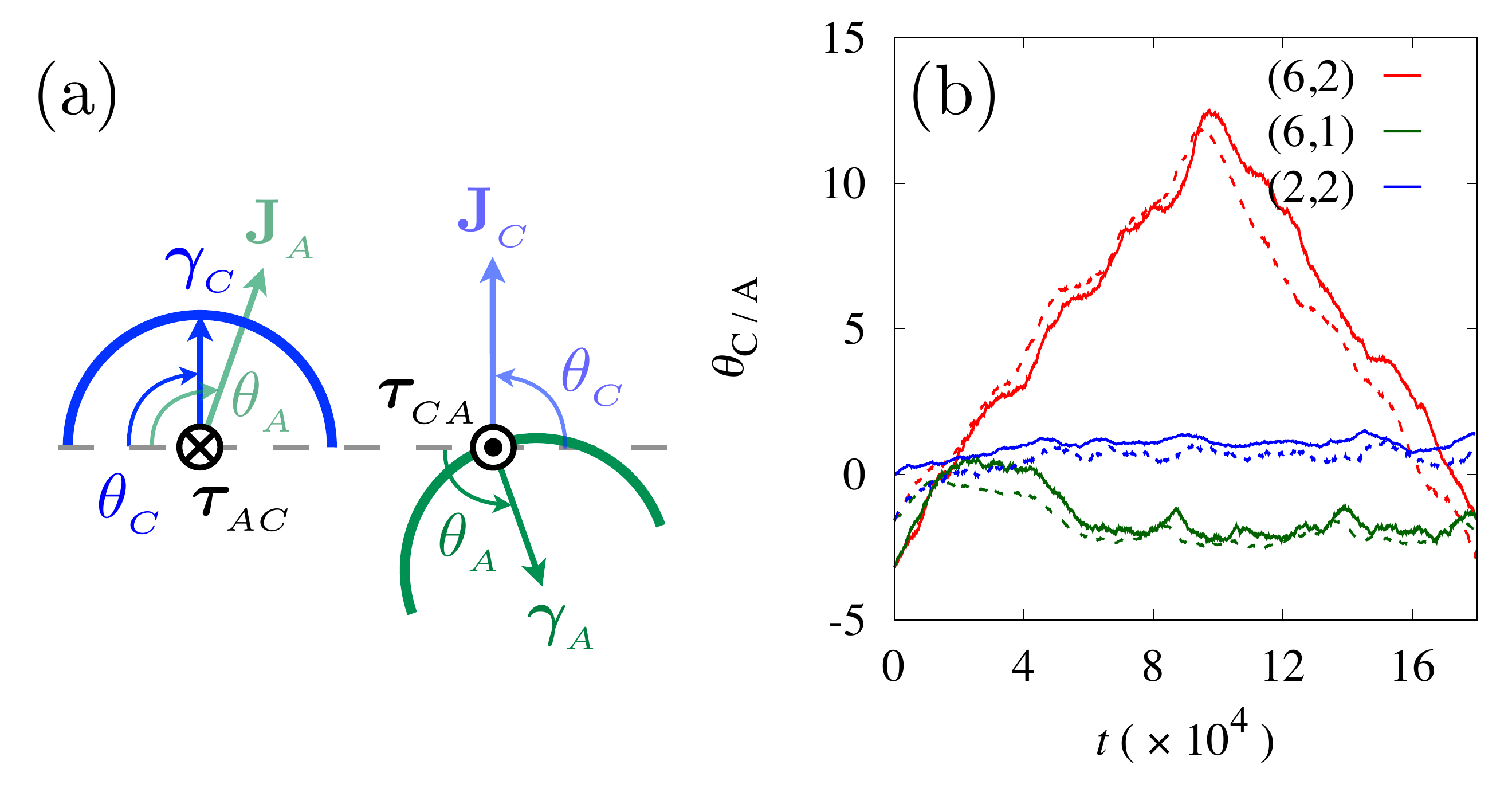}
\caption{\label{fig:fig2} Phase-locking and spontaneous rotations of pinned rotors. (a) An illustration of type~$C$ and
type~$A$ semicircles pinned to the surface (not shown to scale). In the numerics the distance between the pinning points of the semicircles is $5 l_r$. (b) The history of the angles $\theta_{_C}$ (solid lines) and $\theta_{_A}$ (dotted lines) for rotational mobilities given by $(\mu_{_C}^R,\mu_{_A}^R)$. In case $(2,2)$, the solid line indicates $\theta_{_C} + \pi$ instead of $\theta_{_C}$. All the parameters and units of the simulations are specified in Appendix~\ref{app:parameters}.}
\end{figure}

{\em Spontaneous synchronized rotations of pinned rotors.} --- We consider a pair of semicircles, one type~$A$, pinned to the surface at its apex, and one type~$C$, pinned to the surface at its center. They are placed at a distance much larger than $l_r$ so that we
can understand their dynamics in terms of the far-field torques given
in Eq.~\eqref{eq:tau12_sym}. The torques align $\bm\gamma_{_{A/C}}$ with the
current generated by the other circle, which depends only on the
dipole moment, taken to be equal for both. To describe the dynamics we define two angles $\theta_{_C}$ and $\theta_{_A}$ as
the orientation of $\bm\gamma_{_C}$ and $\bm\gamma_{_A}$ with respect
to the horizontal, as represented in Fig.~\ref{fig:fig2}. Note that
$\theta_{_C}$ is defined with a clockwise convention and $\theta_{_A}$
counterclockwise. Neglecting noise, we can write the dynamics of the
angle difference $\theta_{_C}-\theta_{_A}$ in the adiabatic limit as
\begin{align} \label{eq:rotor_pair}
\dot \theta_{_C} -\dot \theta_{_A} &=\frac{1}{\rho_b}\left(\mu_{_C}^R\bm\tau_{_{AC}}-\mu_{_A}^R\bm\tau_{_{CA}}\right)\cdot\mathbf{e}_z \nonumber \\
&=-\frac{\mu_{_C}^R J_{_A} \gamma_{_C}-\mu_{_A}^R J_{_C}\gamma_{_A}}{\rho_b}\sin(\theta_{_C}-\theta_{_A})\;,
\end{align} 
where $\gamma_j \equiv |\bm \gamma_j|$, $J_j \equiv
|\mathbf{J}_j|$. For
$\mu_{_C}^R J_{_A} \gamma_{_C} > \mu_{_A}^R J_{_C} \gamma_{_A}$ the
angles tend to phase-lock at $\theta_C=\theta_A$ while for $\mu_{_C}^R J_{_A} \gamma_{_C} < \mu_{_A}^R J_{_C} \gamma_{_A}$ they phase-lock at $\theta_C=\theta_A+\pi$. Using this we can expand in small deviations from the locking angle
difference. The equation for $\theta_A$ then reduces to
\begin{align} \label{eq:rotorA}
\ddot \theta_{_A} = -\frac{1}{\rho_b}\left| \mu_{_C}^R J_{_A} \gamma_{_C}- \mu_{_A}^R J_{_C} \gamma_{_A}\right|\dot \theta_{_A} \;.
\end{align}
The equation implies that for general parameters $\theta_{_A}$ is
damped and does not rotate persistently along a given
direction. However, if
$\mu_{_C}^R J_{_A} \gamma_{_C} \simeq \mu_{_A}^R J_{_C} \gamma_{_A}$,
the damping is weak and the two rotors persistently counter-rotate
in a spontaneously chosen direction with the same speed and
weak phase-locking interactions. We observe the three types of behavior in 
simulations as shown in the Supplementary Movies \href{http://www.mit.edu/%7Esolon/SI_LR-interaction-Afluid/}{SM1--SM3} and Fig.~\ref{fig:fig2}b.
On the contrary, it is easily seen
that when both rotors are of the same type they always phase-lock,
but do not exhibit persistent rotations. When both are type~$C$ ($A$)
the rotors phase-lock with an angle difference 0 ($\pi$), see SM4 and SM5.

\begin{figure}
\includegraphics[width=0.95\columnwidth]{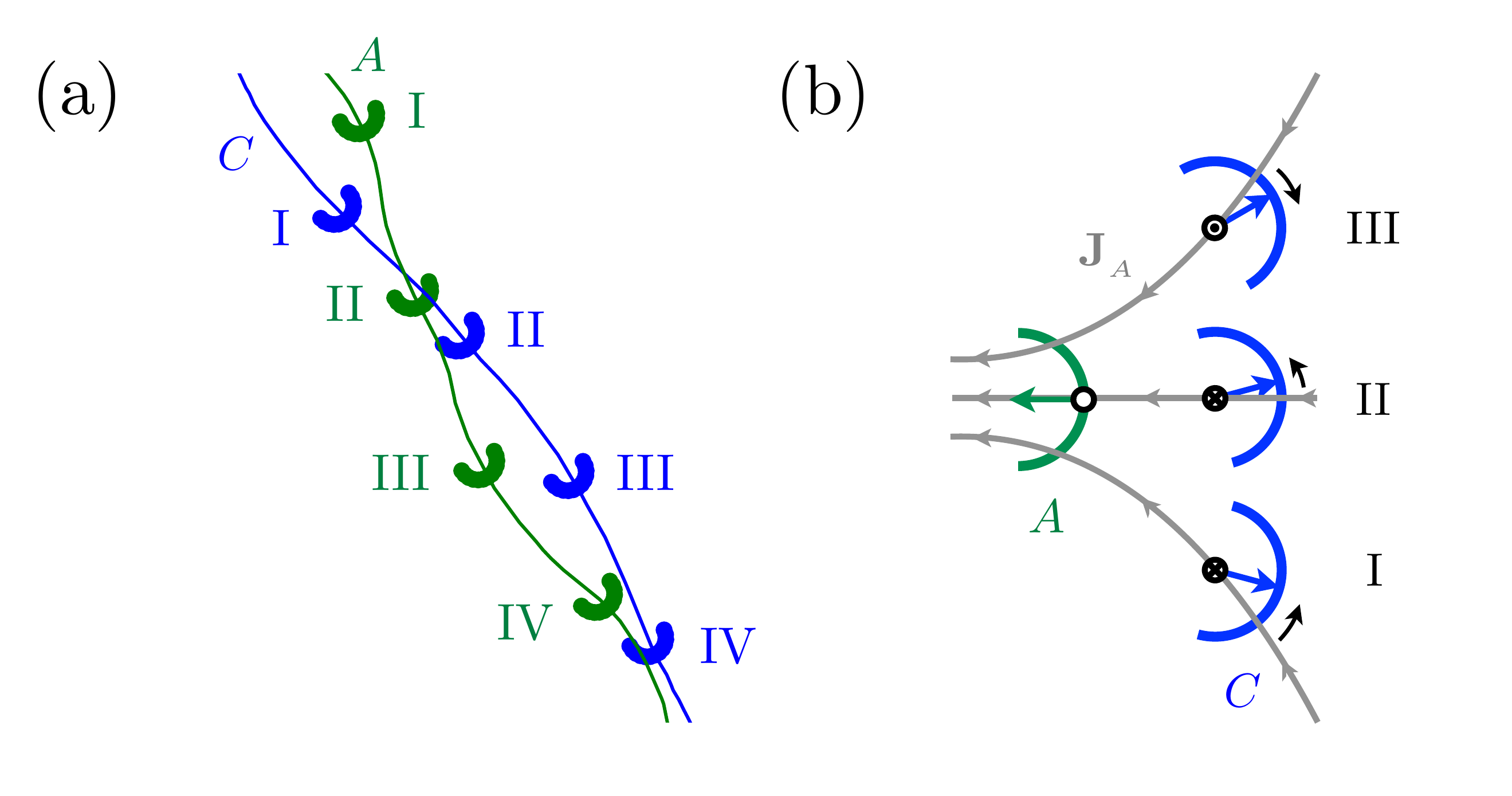}
\caption{\label{fig:fig3} (a) A stroboscopic image of a traveling bound pair obtained from the numerics, with consecutive positions marked by I--IV. All the parameters and units of the simulations are specified in Appendix~\ref{app:parameters}. (b) A schematic illustration of the torques experienced by a type~$C$ body in the current field of a type $A$ body. I, II, and III indicate consecutive relative positions during the snake-like motion. Note that the torque directions are consistent with the resulting motion.}
\end{figure}

{\em Formation of a traveling bound pair.} --- Here we again consider a type~$A$ and type~$C$ pair of semicircles. 
Numerical simulations show that, after a transient regime, the two bodies form a
traveling bound pair with type~$A$ trailing type~$C$ (see
Fig.~\ref{fig:fig3}a and movie SM6). This behavior
can be qualitatively understood in the limit where
$\mu_{_C}^R \gg \mu_{_A}^R$. Then the motion of type~$A$ is, to
leading order, independent of type~$C$ while type~$C$ is strongly
affected by type~$A$. More precisely, the orientation of the type~$C$
body is dictated by its tendency to align $\gamma_{_C}$ with
$\mathbf{J}_{_A}$. This leads to a snake-like motion in front of the
type~$A$ body explained graphically in
Fig.~\ref{fig:fig3}b. Other pairings of semicircles lead to
different phenomena including anti-alignment and the formation of
bound pairs, although in this last case the effect depends on
near-field interactions. These are expected to be less universal than
the far-field interactions and we reserve their study for future work.

In summary, we have explored the long-range interactions occurring
generically between passive bodies placed in an active fluid. We have
shown that, at first order in a multipole expansion, an asymmetric
body generates dipolar currents which decay algebraically in
space. These mediate generic long-range forces and torques between
passive bodies which can be expressed in terms of a few single-body
quantities. Interestingly, the interactions can be tuned by designing
the shape of the bodies which may provide new
routes for designing self-assembling materials. We gave two examples of
dynamical phenomena induced by the far-field interactions. Considering
also near-field effects should reveal many more. Finally, we note that the physics described relies only on the breaking of time-reversal symmetry and a diffusive behavior at large length scale. Therefore, it should be generically present in a broad range of active systems, even including those with mutual interactions between active particles. It would be interesting to check this explicitly using several recent theoretical frameworks~\cite{MaggiSciRep2015,FaragePRE2015,FodorPRL2016}, which have been proposed as approximate descriptions of systems with interacting active particles. Possible relevance of these effects to flocking transitions in shaken granular systems~\cite{KumarNatComm2014} is also of interest.

{\em Acknowledgments.} --- Y.B., N.N., and Y.K. are supported by an I-CORE Program of the Planning and Budgeting Committee of the Israel Science Foundation and an Israel Science Foundation grant. Y.B. is supported in part at the Technion by a fellowship from the Lady Davis Foundation. A.P.S. is supported by the Gordon and Betty Moor foundation through a PLS fellowship. X.X. is supported by Guangdong-Technion Postdoctoral Fellowship.

\onecolumngrid
\appendix

\begin{spacing}{1.5}

\section{Simulation details}

\subsection{Active particles}
\label{app:active_particles}

Molecular dynamics simulations were performed employing an Euler time discretization scheme to integrate the two-dimensional overdamped Langevin dynamics in Eq.~\eqref{eq:langevin} of the main text. At each step of the simulation (from time $t_i$ to $t_{i+1}=t_i + \Delta t$), the position and orientation of an active particle, given by $\left(x,y,\theta\right)$, are updated by
\begin{align} \label{seq:Iteration}
x(t_{i+1}) &= x(t_{i})+\Delta t\,\left[v\cos\theta(t_{i+1})-\mu\,\partial_{x}V+\sqrt{2D_{t}/{\Delta t}}\,\eta_{i}^{x}\right]\;, \nonumber\\
y(t_{i+1}) &= y(t_{i})+\Delta t\,\left[v\sin\theta(t_{i+1})-\mu\,\partial_{y}V+\sqrt{2D_{t}/{\Delta t}}\,\eta_{i}^{y}\right]\;, \nonumber\\
\theta(t_{i+1}) &= \theta(t_{i})+\Delta t\,\sqrt{2D_{t}/{\Delta t}}\,\xi_{i}\;,
\end{align}
where $v$, $\mu$, $D_t$, and $V = V(x,y)$ are defined in the main text, and $\eta_i^{x,y}$, $\xi_i$ are i.i.d. random number sequences with a Gaussian distribution of zero mean and unit variance. The integration time step $\Delta t$ is chosen to provide sufficient accuracy for a given set of parameters, with the highest value employed being $\Delta t = 10^{-2} \, \alpha^{-1}$. At each time step, we first update the orientation $\theta$ and then the spatial coordinates $x$ and $y$. The tumbling dynamics are implemented by choosing a random time interval between tumbles from an exponential distribution with mean $\alpha^{-1}$. If the next tumbling time $t'$ falls between $t_{i}$ and $t_{i+1}$, the interval is divided into two: a regular step takes place between $t_{i}$ and $t'$, after which a new orientation $\theta\left(t_{i+1}\right)$ and the next tumbling time are randomly chosen (the former from a uniform angular distribution). Finally the particle moves on with its new orientation from $t'$ to $t_{i+1}$. For simulations of static bodies, the system is allowed to evolve for a sufficiently long time (adjusted according to the specific parameters in use) before any measurements take place, so that transient effects are removed.

\subsection{Potentials of passive bodies}
\label{app:potentials}

Passive bodies are described by localized potentials. The simplest body employed is a circular bead of radius $R_b$ with a radial harmonic potential
\begin{align}
V_{b}\!\left(\mathbf{r}\right)=\begin{cases}
	\frac{1}{2}\lambda\left(R_b-\left|\mathbf{r}-\mathbf{r}_0\right|\right)^2 & \left|\mathbf{r}-\mathbf{r}_0\right|<R_b\;,\\
			0 & \left|\mathbf{r}-\mathbf{r}_0\right|>R_b\;,
\end{cases}
\end{align}
where $\mathbf{r}_0 = (x_0,y_0)$ is the center of the bead.
The spring constant (or stiffness) $\lambda$ determines the hard-core region, which extends from the center to a radius $R_b-\lambda^{-1}v/\mu$. Only the outer region of the disc surrounding this core can be penetrated by active particles. Due to the circular symmetry, the bead cannot produce any long-range density or current fields on its own.

\begin{figure}
\includegraphics[width=0.7\textwidth]{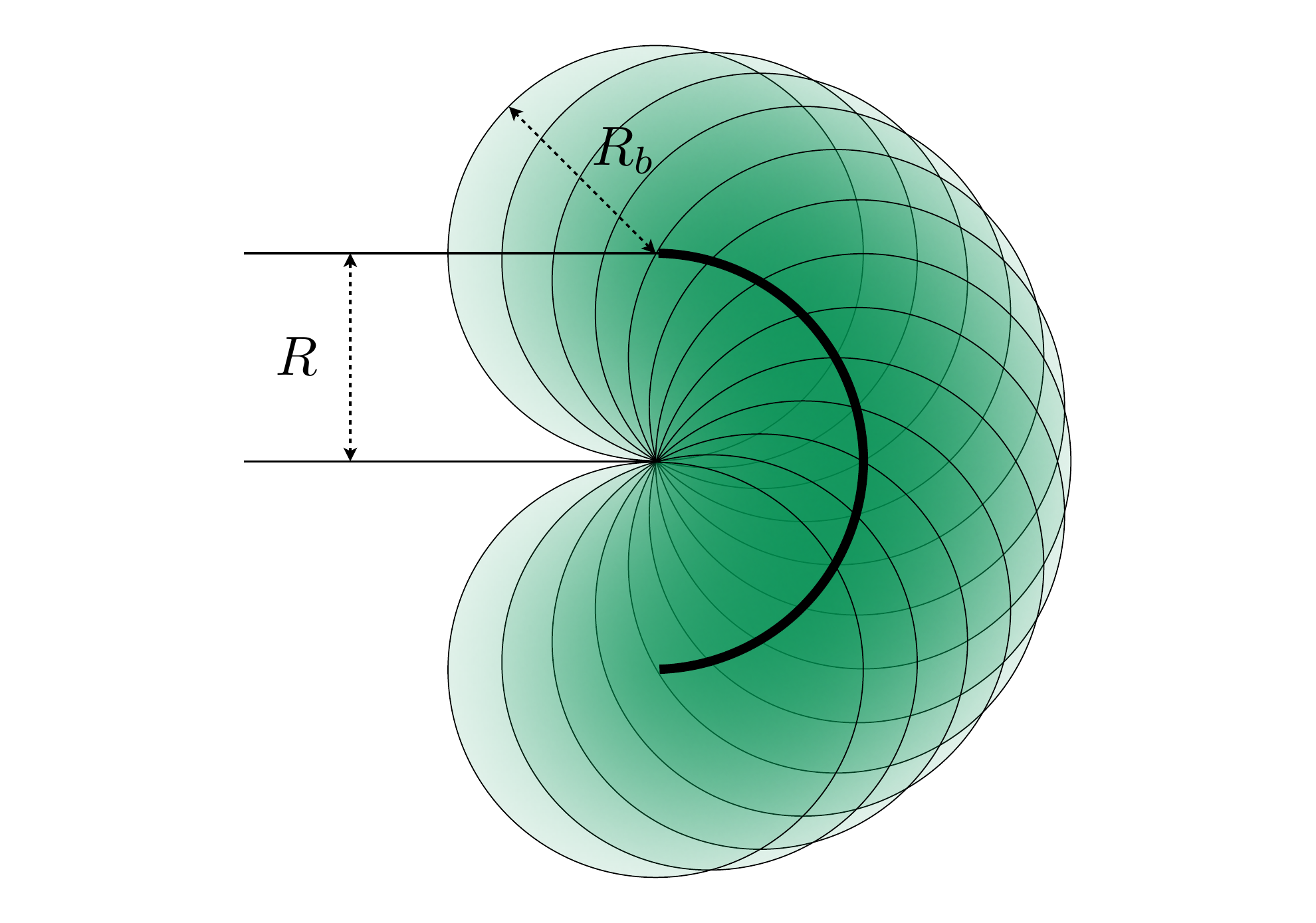}
\caption{\label{fig:figS1} An illustration of a semicircular body of radius $R$ composed of $13$ circular beads of the same radius $R_b = R$. The center of every bead lies on the arc of the semicircle (thick solid line), with any adjacent pair of them separated by equal angular distance.}
\end{figure}

We implemented passive bodies as rigid chains of the beads described above. For asymmetric bodies with nonzero dipole moment, we used
   semicircular bodies composed of identical beads. The beads are arranged so that their centers are placed on the arc of a semicircle of radius $R$. The centers of the two beads at the ends coincide with the tips of the semicircular arc, and all adjacent pairs of beads have equal angular distance from each other. The number of beads $N$ is adjusted to create a sufficient overlap between neighbors, so that the arc of the semicircle is impenetrable to the active particles. A semicircle of radius $R$ composed of $N = 13$ identical beads of the same radius $R_b = R$ is illustrated in Fig.~\ref{fig:figS1}.

In addition, we used rod-like bodies in Fig.~\ref{fig:figS2} that have no dipole moment but non-vanishing quadrupole moment.

\subsection{Dynamics of passive bodies}
\label{app:body_dynamics}

As stated in Eq.~\eqref{eq:body_motion} of the main text, the overdamped dynamics of passive bodies are implemented by first calculating the forces and torques on each of them. At every time step, the total force and torque on body~$j$ are computed by
\begin{align}
\mathbf{F}_{j}(t_i) &= \sum_k\bm\nabla V_{j}\big(\mathbf{r}_k(t_i)\big)\;, \nonumber\\
\bm\tau_{j}(t_i) &= \mathbf{e}_z \cdot \sum_k\left[\mathbf{r}_k(t_i) - \mathbf{R}_{j}(t_i)\right]\times\bm \nabla V_{j}\big(\mathbf{r}_k(t_i)\big)\;,
\end{align}
respectively, where $\mathbf{r}_k$ denotes the position of the $k$-th active particle. The coordinates of body~$j$ are then updated as
\begin{align}
\mathbf{R}_{j}(t_{i+1}) &= \mathbf{R}_{j}(t_i) + \Delta t \, \mu_j^T \,\mathbf{F}_{j}(t_i)\;,\\
\Theta_{j}\left(t_{i+1}\right) &= \Theta_{j}(t_i) + \Delta t \, \mu_j^R \,\bm\tau_{j}(t_i)\;.
\end{align}

\subsection{Simulation parameters} \label{app:parameters}

Here we describe the parameters used in simulations whose results are shown in the figures of the main text and the \href{http://www.mit.edu/%7Esolon/SI_LR-interaction-Afluid/}{Supplementary Movies}. In all simulations listed below, the active fluid is implemented by run-and-tumble particles with $v = \alpha = \mu = 1$ and $D_r = D_t = 0$. Time, lengths and forces are thus measured in units of $\alpha^{-1}$, $l_r=v/\alpha$, and $v/\mu$, respectively.

\begin{itemize}
\item In Fig.~\ref{fig:fig1}, body~$1$ is a semicircle of radius $R = l_r$, which is composed of $127$ circular beads of radius $R_b = l_r/10$ and stiffness $\lambda = 20$. Its normalized dipole moment is measured to be $p_1/\rho_b \simeq 0.58$, headed from the convex side to the concave side of the semicircle. Body~$2$ is a circular bead of radius $l_r/6$ and stiffness $\lambda = 12$. Its normalized response coefficient (corresponding to $\mathbb{R}_2$ in the main text) is measured to be $R_2/\rho_b \simeq 0.025$. If body~$2$ is pinned at a point on the edge, $\bm\gamma_2$ is a vector pointing at the center of the circle from the pinning point, whose magnitude is $0.091$. The mean density of run-and-tumble particles is set to be $\rho_b = 1200$, and the system size is given by $70 \times 70$ with periodic boundaries.
\item In Figs.~\ref{fig:fig2}, \ref{fig:fig3}, and the \href{http://www.mit.edu/%7Esolon/SI_LR-interaction-Afluid/}{Supplementary Movies}, each semicircle has radius $R = l_r$ and is composed of $13$ identical beads of radius $R_b = l_r$ and stiffness $\lambda = 2$. Type~$C$ semicircles rotate about the center of the semicircle, and type~$A$ semicircles rotate about the center of the middle bead. In Fig.~\ref{fig:fig3}, the translational mobilities of both semicircles are fixed at $\mu^T_{_C} = \mu^T_{_A} = 1$. The mean density of run-and-tumble particles is set to be $\rho_b = 800$, and the system size is given by $30 \times 30$ with periodic boundaries.
\end{itemize}

\section{Derivation of far-field density and current}
\label{app:density_current}

% Eq. (9): eq:dipole_density
% Eq. (10): eq:p_force
% Eq. (3): eq:fokker-planck
% Eq. (4): eq:marginal
In this section, we present a detailed derivation of the far-field active particle density $\rho_j = \rho_j(\mathbf{r})$ created by a static body~$j$, whose leading order behavior is given in Eqs.~\eqref{eq:dipole_density} and \eqref{eq:dipole_moment} of the main text. We start from the steady-state Fokker--Planck equation for the distribution $P_j = P_j(\mathbf{r},\theta)$ of active particles
\begin{align} \label{seq:fokker-planck_rep}
	0 = -\bm\nabla\cdot\left(v \mathbf{e}_\theta-\mu\bm\nabla V_j -D_t\bm\nabla\right)P_j - \alpha P_j
	+ \frac{\alpha}{2\pi}\int d\theta\, P_j + D_r\partial_\theta^2 P_j\;,
\end{align}
which corresponds to the Langevin dynamics given by Eq.~\eqref{eq:langevin} of the main text. Next, we introduce the marginal distributions
\begin{align}
	\mathbf{m}_j^{(n)}(\mathbf{r}) \equiv \int d\theta\, P_j(\mathbf{r},\theta)\begin{bmatrix}
		\cos (n\theta)\\
		\sin (n\theta)	
	\end{bmatrix}
\end{align}
for integers $n \ge 0$, with $\mathbf{m}_j^{(0)} = (\rho_j,0)^{T}$. Multiplying both sides of Eq.~\eqref{seq:fokker-planck_rep} by $\cos (n\theta)$ or $\sin (n\theta)$ and integrating over $\theta$, we obtain the hierarchical relations
\begin{align} \label{seq:zeroth}
	0 &= -\bm\nabla\cdot \left(v\mathbf{m}_j^{(1)}-\rho_j\mu\bm\nabla V_j-D_t\bm\nabla\rho_j\right)\;, \\
	0 &= -(\alpha+D_r n^2)(1-\mathbb{M}_j^{(n)})\mathbf{m}_j^{(n)}-\frac{v}{2}\left(\mathbb{D}\mathbf{m}_j^{(n-1)}-\mathbb{D}^\dagger\mathbf{m}_j^{(n+1)}\right)\;, \label{seq:nth}
\end{align}
where the second equation holds for integers $n \ge 1$, and we have defined the linear operators
\begin{align}
\mathbb{M}_j^{(n)} &\equiv \frac{1}{\alpha+D_r n^2}\left[\mu \bm \nabla \cdot (\bm \nabla V_j)+D_t\nabla^2\right] \nonumber\\
	 &= \frac{1}{\alpha+D_r n^2}\left[\mu (\nabla^2 V_j)+\bm (\nabla V_j) \cdot \bm \nabla+D_t\nabla^2\right]\;,\label{seq:M_def}\\
	\mathbb{D} &\equiv \begin{bmatrix*}[r]
		\partial_x & -\partial_y \\
		\partial_y & \partial_x
	\end{bmatrix*}\;, \quad
	\mathbb{D}^\dagger = \begin{bmatrix*}[r]
		-\partial_x & -\partial_y \\
		\partial_y & -\partial_x
	\end{bmatrix*}\;. \label{seq:D_def}
\end{align}
The operators $\mathbb{D}$ and $\mathbb{D}^\dagger$ form a conjugate pair satisfying
\begin{align} \label{seq:D_Ddagger}
	\mathbb{D}\mathbb{D}^\dagger = \mathbb{D}^\dagger\mathbb{D} = -\nabla^2\;.
\end{align}
We also note that solving Eq.~\eqref{seq:nth} for $\mathbf{m}_j^{(n)}$ gives
\begin{align} \label{seq:m_hier}
	\mathbf{m}_j^{(n)} &= -\frac{v}{2(\alpha+D_r n^2)}\Big(1-\mathbb{M}_j^{(n)}\Big)^{-1}\Big(\mathbb{D}\mathbf{m}_j^{(n-1)}-\mathbb{D}^\dagger\mathbf{m}_j^{(n+1)}\Big) \nonumber\\
	&= -\frac{v}{2(\alpha+D_r n^2)}\sum_{k=0}^\infty\Big(\mathbb{M}_j^{(n)}\Big)^k\Big(\mathbb{D}\mathbf{m}_j^{(n-1)}-\mathbb{D}^\dagger\mathbf{m}_j^{(n+1)}\Big)\;,
\end{align}
where the second equation is obtained by a formal expansion.

For $n = 1$ and $n = 2$, Eq.~\eqref{seq:nth} can be rewritten as
\begin{align}
\mathbf{m}_j^{(1)} &= \mathbb{M}_j^{(1)}\mathbf{m}_j^{(1)} - \frac{l_r}{2}\left(\bm\nabla\rho_j + \mathbb{D}^\dagger\mathbf{m}_j^{(2)}\right)\;, \label{seq:m1}\\
\mathbf{m}_j^{(2)} &= \mathbb{M}_j^{(2)}\mathbf{m}_j^{(2)} - \frac{l_r}{2}\frac{\alpha+D_r}{\alpha+4D_r}\left(\mathbb{D}\mathbf{m}_j^{(1)} - \mathbb{D}^\dagger\mathbf{m}_j^{(3)}\right)\;, \label{seq:m2}
\end{align}
with $l_r \equiv v/(\alpha+D_r)$ representing the run length of each active particle. Applying $-v \bm \nabla \cdot$ to both sides of Eq.~\eqref{seq:m1} and expressing $\mathbb{M}_j^{(1)}$ by Eq.~\eqref{seq:M_def}, we obtain
\begin{align}
-\bm\nabla\cdot\left(v\mathbf{m}_j^{(1)}\right) &= -\mu l_r \sum_{a,\,b} \partial_a\partial_b \Big(\partial_a V_j\Big)\left(\mathbf{m}_j^{(1)}\cdot\mathbf{e}_b\right) - D_t l_r \nabla^2 \bm\nabla\cdot \mathbf{m}_j^{(1)}\nonumber\\
&\quad + \frac{v l_r}{2}\left(\nabla^2\rho_j + \bm\nabla\cdot \mathbb{D}^\dagger \mathbf{m}_j^{(2)}\right)\;,
\end{align}
where the indices $a$ and $b$ run over the Cartesian coordinates $\{x,\,y\}$, and $\mathbf{e}_b$ denotes the unit vector in the $b$-direction. This equation can be further expanded by replacing $\mathbf{m}_j^{(2)}$ with Eq.~\eqref{seq:m2}, and then using Eq.~\eqref{seq:M_def} with $n = 2$ and Eq.~\eqref{seq:D_Ddagger}. One obtains
\begin{align} \label{seq:div_m1}
-\bm\nabla\cdot\left(v\mathbf{m}_j^{(1)}\right) &= \frac{v l_r}{2}\nabla^2\rho_j - \mu l_r \sum_{a,\,b} \partial_a\partial_b \Big(\partial_a V_j\Big)\left(\mathbf{m}_j^{(1)}\cdot\mathbf{e}_b\right) \nonumber\\
&\quad + \sum_{a,\,b,\,c} \partial_a\partial_b\partial_c (\mathbb{H}_j)_{abc}\;,
\end{align}
with $a$, $b$, and $c$ running over $\{x,\,y\}$, and a rank-$3$ tensor $\mathbb{H}_j$ satisfying
\begin{align} \label{seq:H_def}
\sum_{a,\,b,\,c} \partial_a\partial_b\partial_c (\mathbb{H}_j)_{abc} &= - D_t l_r \nabla^2 \bm\nabla\cdot \mathbf{m}_j^{(1)} - \frac{v l_r^2}{4}\frac{\alpha+D_r}{\alpha+4D_r}\left[\nabla^2\bm\nabla\cdot\mathbf{m}_j^{(1)}+\bm\nabla\cdot\left(\mathbb{D}^\dagger\right)^2\mathbf{m}_j^{(3)}\right] \nonumber\\
&\quad + \frac{l_r^2}{2}\frac{\alpha+D_r}{\alpha+4D_r}\bm\nabla\cdot\mathbb{D}^\dagger\sum_a \partial_a\left[\mu(\partial_a V_j) + D_t\partial_a\right]\mathbf{m}_j^{(2)}\;.
\end{align}
Using Eq.~\eqref{seq:div_m1} to replace $\mathbf{m}_j^{(1)}$ in Eq.~\eqref{seq:zeroth} leads to the two-dimensional Poisson equation
\begin{align} \label{seq:poisson}
0 = D_\text{eff}\nabla^2\rho_j + \mu\bm\nabla\cdot\left(\rho_j\bm\nabla V_j\right) - \mu l_r \sum_{a,\,b}\partial_a\partial_b\Big(\partial_a V_j\Big)\left(\mathbf{m}_j^{(1)}\cdot\mathbf{e}_b\right) + \sum_{a,\,b,\,c} \partial_a\partial_b\partial_c (\mathbb{H}_j)_{abc}\;,
\end{align}
where $D_\text{eff} \equiv D_t + v l_r/2$ is the effective diffusion constant. We note that the equation is consistent with Eq.~\eqref{eq:poisson} of the main text, in which the rank-$2$ tensor $\mathbb{G}_j$ is given by
\begin{align} \label{seq:G_def}
(\mathbb{G}_j)_{ab} \equiv -\mu l_r \Big(\partial_a V_j\Big)\left(\mathbf{m}_j^{(1)}\cdot\mathbf{e}_b\right) + \sum_c \partial_c (\mathbb{H}_j)_{abc}\;.
\end{align}

Applying the method of Green's functions, the solution to Eq.~\eqref{seq:poisson} is obtained as
\begin{align} \label{seq:rho_green}
\rho_j(\mathbf{r}) &= \rho_b -\frac{\beta_\text{eff}}{2\pi}\int d^2\mathbf{r}'\, \ln |\mathbf{r}-\mathbf{r}'|\nonumber\\
&\quad \times\left[\bm\nabla'\cdot (\rho_j\bm\nabla'V_j) - l_r \sum_{a,\,b}\partial'_a\partial'_b\Big(\partial'_a V_j\Big)\left(\mathbf{m}_j^{(1)}\cdot\mathbf{e}_b\right) + \frac{1}{\mu}\sum_{a,\,b,\,c} \partial'_a\partial'_b\partial'_c (\mathbb{H}_j)_{abc}\right]\;,
\end{align}
where $\beta_\text{eff} \equiv \mu/D_\text{eff}$ denotes the effective inverse temperature, and primed derivatives are with respect to $\mathbf{r}'$. In the far-field regime, i.e. when $r \equiv |\mathbf{r}|$ is greater than any other microscopic length scale, the integral in Eq.~\eqref{seq:rho_green} can be approximated by a multipole expansion
\begin{align} \label{seq:rho_multipole}
\rho_j(\mathbf{r}) = \rho_b + \frac{\beta_\text{eff}}{2\pi}\,\left(\frac{\mathbf{r}\cdot\mathbf{p}_j}{r^2} + \frac{\mathbf{r}\cdot\mathbb{Q}_j\mathbf{r}}{2r^4}\right) + O(r^{-3})\;.
\end{align}
Here the dipole moment $\mathbf{p}_j$ is given by
\begin{align} \label{seq:dipole_moment}
\mathbf{p}_j = -\int d^2\mathbf{r}'\,\rho_j\bm\nabla'V\;,
\end{align}
and the quadrupole moment $\mathbb{Q}_j$ satisfies
\begin{align} \label{seq:quadrupole_moment}
(\mathbb{Q}_j)_{ab} &= -2\int d^2\mathbf{r}'\, \Bigg\{\rho_j(r'_a\,\partial'_b V_j + r'_b\,\partial'_a V_j) - l_r \left[\Big(\partial'_a V_j\Big)\left(\mathbf{m}_j^{(1)}\cdot\mathbf{e}_b\right) + \Big(\partial'_b V_j\Big)\left(\mathbf{m}_j^{(1)}\cdot\mathbf{e}_a\right)\right] \nonumber\\
&\quad -\delta_{ab}\left[\rho_j\mathbf{r}'\cdot\bm\nabla'V_j - l_r\Big(\bm\nabla' V_j\Big)\cdot\mathbf{m}_j^{(1)}\right]\Bigg\}\;.
\end{align}
Note that at dipole order there are no contributions from tensor $\mathbb{G}_j$ defined in Eq.~\eqref{seq:G_def}, and that at quadrupole order there are no contributions from tensor $\mathbb{H}_j$ expressed in Eq.~\eqref{seq:H_def}. In principle, one can calculate higher-order multipoles from the contributions of $\mathbb{H}_j$, which are beyond the scope of this work.

The results are consistent with Eqs.~\eqref{eq:dipole_density} and \eqref{eq:dipole_moment} of the main text. We also note that Eqs.~\eqref{seq:dipole_moment} and \eqref{seq:quadrupole_moment} are {\em exact}; they do not rely on any further assumptions besides the far-field limit. Moreover, based on the equations, $\mathbf{p}_j$ and $\mathbb{Q}_j$ can be measured numerically using the marginal distributions $\rho_j$ and $\mathbf{m}_j^{(1)}$ of active particles on body~$j$. Both can also be estimated from the far-field behavior using Eq.~\eqref{seq:rho_multipole}.

We now turn to the current density of active particles $\mathbf{J}_j = \mathbf{J}_j(\mathbf{r})$ generated by static body~$j$. Since Eq.~\eqref{seq:zeroth} is a continuity equation $\bm\nabla\cdot\mathbf{J}_j = 0$, the current density is given by
\begin{align} \label{seq:current_field}
\mathbf{J}_j = v\mathbf{m}_j^{(1)} - \rho_j\mu\bm\nabla V_j - D_t\bm\nabla\rho_j
= -D_\text{eff}\bm\nabla\rho_j - \rho_j\mu\bm\nabla V_j + v\mathbb{M}_j^{(1)}\mathbf{m}_j^{(1)}-\frac{v l_r}{2}\mathbb{D}^\dagger\mathbf{m}_j^{(2)}\;,
\end{align}
where the second equation is obtained by replacing $\mathbf{m}_j^{(1)}$ with Eq.~\eqref{seq:m1}. We now show that the last two terms in the equation are of order $r^{-3}$ in the far-field. To see this, note that the recursive structure of Eq.~\eqref{seq:m_hier} implies $\mathbf{m}_j^{(n)} \sim \partial^n \rho_j + O(\partial^{n+2}\rho_j)$, with $\partial$ denoting a generic spatial derivative. In the far field we have $\mathbb{M}_j^{(1)} \sim \partial^2$ and $\mathbb{D}^\dagger \sim \partial$, thus we are left with
\begin{align} \label{seq:current_exp}
\mathbf{J}_j(\mathbf{r}) &= -D_\text{eff}\bm\nabla\rho_j + O(\partial^3\rho_j)\nonumber\\
&= -\frac{\mu}{2\pi}\left[\frac{\mathbf{p}_j}{r^2} - \frac{2(\mathbf{r}\cdot\mathbf{p}_j)\mathbf{r}}{r^4}\right]
-\frac{\mu}{2\pi}\left[\frac{\mathbb{Q}_j\mathbf{r}}{r^4} - \frac{2(\mathbf{r}\cdot\mathbb{Q}_j\mathbf{r})\mathbf{r}}{r^6}\right] + O(r^{-4}) \;,
\end{align}
which is dominated by the diffusive component. The dipole component of this expression is given in Eq.~\eqref{eq:dipole_current} of the main text. Moreover, by integrating both sides of Eq.~\eqref{seq:current_field} over the entire space, we obtain
\begin{align}
	\int d^2\mathbf{r}\,\mathbf{J}_j = -\mu\int d^2\mathbf{r}\,\rho_j\bm\nabla V_j = \mu \mathbf{p}_j\;,
\end{align}
which is an exact current--force relation for this class of model, derived for more general cases in \cite{NikolaPRL2016}.

\section{Interactions between distant bodies}
\label{app:interactions}

We consider a pair of static bodies in an active fluid of bulk density $\rho_b$. Their interactions with active particles are described by potentials $V_1$ and $V_2$. Body~$2$ is positioned at the origin ($\mathbf{R}_2 = 0$), and body~$1$ is at a distant location $\mathbf{R}_1 = \mathbf{r}_{12}$. Extending Eq.~\eqref{seq:rho_green} to the two-body case, the marginal distributions $\rho = \rho(\mathbf{r})$ and $\mathbf{m}^{(n)} = \mathbf{m}^{(n)}(\mathbf{r})$ of the active particles satisfy
\begin{align} \label{seq:rho_green_2}
\rho(\mathbf{r}) &= \rho_b - \frac{\beta_\text{eff}}{2\pi}\int d^2\mathbf{r}'\, \ln |\mathbf{r}-\mathbf{r}'|\nonumber\\
&\quad \times\left[\bm\nabla'\cdot (\rho\bm\nabla'V) - l_r \sum_{a,\,b}\partial'_a\partial'_b\Big(\partial'_a V\Big)\Big(\mathbf{m}^{(1)}\cdot\mathbf{e}_b\Big) + \frac{1}{\mu}\sum_{a,\,b,\,c} \partial'_a\partial'_b\partial'_c (\mathbb{H})_{abc}\right]\;,
\end{align}
where $V = V_1 + V_2$, and $\mathbb{H}$ satisfies
\begin{align} \label{seq:H_2body_def}
\sum_{a,\,b,\,c} \partial_a\partial_b\partial_c (\mathbb{H})_{abc} &= - D_t l_r \nabla^2 \bm\nabla\cdot \mathbf{m}^{(1)} - \frac{v l_r^2}{4}\frac{\alpha+D_r}{\alpha+4D_r}\left[\nabla^2\bm\nabla\cdot\mathbf{m}^{(1)}+\bm\nabla\cdot\left(\mathbb{D}^\dagger\right)^2\mathbf{m}^{(3)}\right] \nonumber\\
&\quad + \frac{l_r^2}{2}\frac{\alpha+D_r}{\alpha+4D_r}\bm\nabla\cdot\mathbb{D}^\dagger\sum_a \partial_a\left[\mu(\partial_a V) + D_t\partial_a\right]\mathbf{m}^{(2)}\;.
\end{align}
This relation for $\mathbb{H}$ is similar to Eqs.~\eqref{seq:H_def} for $\mathbb{H}_j$, but the index~$j$ is dropped in the former to address the full two-body problem. In order to obtain the force $\mathbf{F}_{12}$ applied by body~$1$ on body~$2$, we focus on $\mathbf{r}$ in the vicinity of body~$2$. To this end, we divide the spatial integral into two domains: $\Omega_1$ covers the domain of body~$1$, and $\Omega_1^\mathrm{c}$ covers the rest of the space. Then the integral can be decomposed as
\begin{align} \label{seq:int_1_plus_int_2}
\rho(\mathbf{r}) &= \rho_b - \frac{\beta_\text{eff}}{2\pi}\int_{\Omega_1} d^2\mathbf{r}'\, \ln |\mathbf{r}-\mathbf{r}'|\nonumber\\
&\quad \times\left[\bm\nabla'\cdot \Big(\rho\bm\nabla'V_1\Big) - l_r \sum_{a,\,b}\partial'_a\partial'_b\Big(\partial'_a V_1\Big)\Big(\mathbf{m}^{(1)}\cdot\mathbf{e}_b\Big) + \frac{1}{\mu}\sum_{a,\,b,\,c} \partial'_a\partial'_b\partial'_c (\mathbb{H})_{abc}\right]\nonumber\\
&-\frac{\beta_\text{eff}}{2\pi}\int_{\Omega_1^\mathrm{c}} d^2\mathbf{r}'\, \ln |\mathbf{r}-\mathbf{r}'|\nonumber\\
&\quad \times\left[\bm\nabla'\cdot \Big(\rho\bm\nabla'V_2\Big) - l_r \sum_{a,\,b}\partial'_a\partial'_b\Big(\partial'_a V_2\Big)\Big(\mathbf{m}^{(1)}\cdot\mathbf{e}_b\Big) + \frac{1}{\mu}\sum_{a,\,b,\,c} \partial'_a\partial'_b\partial'_c (\mathbb{H})_{abc}\right] \;.
\end{align}
For large $r_{12}$, the contribution from $\Omega_1$ is always a far-field effect and can be expressed by a multipole expansion
\begin{align} \label{seq:integral_1_multipole_exp}
&- \frac{\beta_\text{eff}}{2\pi}\int_{\Omega_1} d^2\mathbf{r}'\, \ln |\mathbf{r}-\mathbf{r}'|\left[\bm\nabla'\cdot \Big(\rho\bm\nabla'V_1\Big) - l_r \sum_{a,\,b}\partial'_a\partial'_b\Big(\partial'_a V_1\Big)\Big(\mathbf{m}^{(1)}\cdot\mathbf{e}_b\Big) + \frac{1}{\mu}\sum_{a,\,b,\,c} \partial'_a\partial'_b\partial'_c (\mathbb{H})_{abc}\right] \nonumber\\
&\quad = \frac{\beta_\text{eff}}{2\pi}\left[\frac{(\mathbf{r}_{12}+\mathbf{r})\cdot\tilde{\mathbf{p}}_1}{|\mathbf{r}_{12}+\mathbf{r}|^2} + \frac{\mathbf{r}_{12}\cdot\tilde{\mathbb{Q}}_1\mathbf{r}_{12}}{2r_{12}^4}\right] + O\!\left(r_{12}^{-3}\right)\nonumber\\
&\quad = \frac{\beta_\text{eff}}{2\pi}\left[\frac{\mathbf{r}_{12}\cdot\tilde{\mathbf{p}}_1}{r_{12}^2}+\frac{\mathbf{r}\cdot\tilde{\mathbf{p}}_1}{r_{12}^2}- \frac{2(\mathbf{r}\cdot\mathbf{r}_{12})(\mathbf{r}_{12}\cdot\tilde{\mathbf{p}}_1)}{r_{12}^4} + \frac{\mathbf{r}_{12}\cdot\tilde{\mathbb{Q}}_1\mathbf{r}_{12}}{2r_{12}^4}\right] + O\!\left(r_{12}^{-3}\right)\;,
\end{align}
where $\tilde{\mathbf{p}}_j$ and $\tilde{\mathbb{Q}}_j$ are the modified dipole and quadrupole moments of body~$j$ given by
\begin{align}
\tilde{\mathbf{p}}_j &= -\int d^2\mathbf{r}'\,\rho\bm\nabla'V_j\;, \label{seq:dipole_moment_2}\\
\left(\tilde{\mathbb{Q}}_j\right)_{ab} &= -2\int d^2\mathbf{r}'\, \Bigg\{\rho\,(r'_a\,\partial'_b V_j + r'_b\,\partial'_a V_j) - l_r \left[\big(\partial'_a V_j\big)\big(\mathbf{m}^{(1)}\cdot\mathbf{e}_b\big) + \big(\partial'_b V_j\big)\big(\mathbf{m}^{(1)}\cdot\mathbf{e}_a\big)\right] \nonumber\\
&\quad -\delta_{ab}\left[\rho\,\mathbf{r}'\cdot\bm\nabla'V_j - l_r(\bm\nabla' V_j)\cdot\mathbf{m}^{(1)}\right]\Bigg\}\;, \label{seq:quadrupole_moment_2}
\end{align}
respectively. Note that, in contrast to $\mathbf{p}_j$ and $\mathbb{Q}_j$ defined for isolated body~$j$ in Eqs.~\eqref{seq:dipole_moment} and \eqref{seq:quadrupole_moment}, $\tilde{\mathbf{p}}_j$ and $\tilde{\mathbb{Q}}_j$ also take into account the influence of the other body through $\rho$ and $\mathbf{m}^{(1)}$, which are solutions of the full two-body problem.

Using Eq.~\eqref{seq:integral_1_multipole_exp} in Eq.~\eqref{seq:int_1_plus_int_2}, we obtain
\begin{align} \label{seq:rho_green_2_exp_0}
\rho(\mathbf{r}) &= -\frac{\beta_\text{eff}}{2\pi}\int_{\Omega_1^\mathrm{c}} d^2\mathbf{r}'\, \ln |\mathbf{r}-\mathbf{r}'|\nonumber\\
&\quad\times\left[\bm\nabla'\cdot \Big(\rho\bm\nabla'V_2\Big) - l_r \sum_{a,\,b}\partial'_a\partial'_b\Big(\partial'_a V_2\Big)\Big(\mathbf{m}^{(1)}\cdot\mathbf{e}_b\Big) + \frac{1}{\mu}\sum_{a,\,b,\,c} \partial'_a\partial'_b\partial'_c (\mathbb{H})_{abc}\right] \nonumber\\
&\quad + \tilde{\rho}_b - \frac{\mathbf{r}\cdot\tilde{\mathbf{J}}_b}{D_\text{eff}} + O\!\left(r_{12}^{-3}\right)\;,
\end{align}
where we introduced the shorthand notations
\begin{align} \label{seq:mod_rho_b_J_b}
\tilde{\rho}_b = \rho_b + \frac{\beta_\text{eff}}{2\pi}\left[\frac{\mathbf{r}_{12}\cdot\tilde{\mathbf{p}}_1}{r_{12}^2} + \frac{\mathbf{r}_{12}\cdot\tilde{\mathbb{Q}}_1\mathbf{r}_{12}}{2r_{12}^4}\right]\;, \quad
\tilde{\mathbf{J}}_b = -\frac{\mu}{2\pi}\left[\frac{\tilde{\mathbf{p}}_1}{r_{12}^2} - \frac{2(\mathbf{r}_{12}\cdot\tilde{\mathbf{p}}_1)\mathbf{r}_{12}}{r_{12}^4}\right]\;.
\end{align}
We can further simplify the integral in Eq.~\eqref{seq:rho_green_2_exp_0} by noting that, even if we add $\Omega_1$ to the range of the integral, the contribution of the added domain is of order $r_{12}^{-3}$ due to vanishing $V_2$ and the three spatial derivatives in front of $\mathbb{H}$. Thus, up to order $r_{12}^{-2}$, the integral over $\Omega_1^\mathrm{c}$ can be safely replaced with the one over the entire space
\begin{align} \label{seq:rho_green_2_exp}
\rho(\mathbf{r}) &= -\frac{\beta_\text{eff}}{2\pi}\int d^2\mathbf{r}'\, \ln |\mathbf{r}-\mathbf{r}'|\nonumber\\
&\quad\times\left[\bm\nabla'\cdot \Big(\rho\bm\nabla'V_2\Big) - l_r \sum_{a,\,b}\partial'_a\partial'_b\Big(\partial'_a V_2\Big)\Big(\mathbf{m}^{(1)}\cdot\mathbf{e}_b\Big) + \frac{1}{\mu}\sum_{a,\,b,\,c} \partial'_a\partial'_b\partial'_c (\tilde{\mathbb{H}}_2)_{abc}\right] \nonumber\\
&\quad + \tilde{\rho}_b - \frac{\mathbf{r}\cdot\tilde{\mathbf{J}}_b}{D_\text{eff}} + O\!\left(r_{12}^{-3}\right)\;,
\end{align}
where $\tilde{\mathbb{H}}_2$ satisfies
\begin{align} \label{seq:Htilde_2body_def}
\sum_{a,\,b,\,c} \partial_a\partial_b\partial_c (\tilde{\mathbb{H}}_2)_{abc} &= - D_t l_r \nabla^2 \bm\nabla\cdot \mathbf{m}^{(1)} - \frac{v l_r^2}{4}\frac{\alpha+D_r}{\alpha+4D_r}\left[\nabla^2\bm\nabla\cdot\mathbf{m}^{(1)}+\bm\nabla\cdot\left(\mathbb{D}^\dagger\right)^2\mathbf{m}^{(3)}\right] \nonumber\\
&\quad + \frac{l_r^2}{2}\frac{\alpha+D_r}{\alpha+4D_r}\bm\nabla\cdot\mathbb{D}^\dagger\sum_a \partial_a\left[\mu(\partial_a V_2) + D_t\partial_a\right]\mathbf{m}^{(2)}\;,
\end{align}
which is similar to Eq.~\eqref{seq:H_2body_def} except that $V = V_1+V_2$ is replaced with $V_2$.

We note that Eq.~\eqref{seq:rho_green_2_exp} is similar to Eq.~\eqref{seq:rho_green} for a single passive body, the only differences being the modification of $\rho_b$ to $\tilde{\rho}_b$ and a global diffusive current imposed by
\begin{align}
	-\lim_{r\to\infty} D_\text{eff}\bm\nabla \rho = \tilde{\mathbf{J}}_b\;,
\end{align}
which is absent in Eq.~\eqref{seq:rho_green}. Thus Eq.~\eqref{seq:rho_green_2_exp} can be interpreted as the steady-state condition for the distribution of active particles around isolated body~$2$, with the constraints that the global mean density of the active particles is given by $\tilde{\rho}_b$ and that a boundary-driven diffusive current $\tilde{\mathbf{J}}_b$ flows through the system. Since the active particles are mutually independent, the shape of their steady-state distribution is determined by the normalized current $\mathbf{u} \equiv \tilde{\mathbf{J}}_b/\tilde{\rho}_b$, which can also be regarded as the effective velocity of each active particle carrying the current. Based on these considerations, the solution to Eq.~\eqref{seq:rho_green_2_exp} can be written as
\begin{align}
	\rho(\mathbf{r}) = \frac{\tilde{\rho}_b}{\rho_b}\rho_2^{(\mathbf{u})}(\mathbf{r})\;,
	\quad \mathbf{m}^{(n)}(\mathbf{r}) = \frac{\tilde{\rho}_b}{\rho_b}\mathbf{m}_2^{(n,\mathbf{u})}(\mathbf{r}) \text{ for integers $n\ge 1$,} 
\end{align}
where $\rho_2^{(\mathbf{u})}$ and $\mathbf{m}_2^{(n,\mathbf{u})}$ are associated with the steady-state distribution of active particles around body~$2$ when the global mean density and the effective velocity are given by $\rho_b$ and $\mathbf{u}$, respectively. Note that $\rho_2 = \rho_2^{(\mathbf{0})}$ and $\mathbf{m}_2^{(n)} = \mathbf{m}_2^{(n,\mathbf{0})}$ solve the single-body problem~\eqref{seq:rho_green} for $j = 2$. Since $u \equiv |\mathbf{u}| \sim |\tilde{\mathbf{J}}_b| \sim r_{12}^{-2}$, for large $r_{12}$ one can use linear approximations
\begin{align}
\rho(\mathbf{r}) &= \frac{\tilde{\rho}_b}{\rho_b}\rho_2 + \frac{\tilde{\rho}_b}{\rho_b}\mathbf{u}\cdot \left.\bm\nabla_\mathbf{u}\rho_2^{(\mathbf{u})}\right|_{\mathbf{u}=\mathbf{0}} + O\!\left(u^2,r_{12}^{-3}\right) \nonumber\\
&= \frac{\tilde{\rho}_b}{\rho_b}\rho_2 + \frac{1}{\rho_b}\tilde{\mathbf{J}}_b\cdot \left.\bm\nabla_\mathbf{u}\rho_2^{(\mathbf{u})}\right|_{\mathbf{u}=\mathbf{0}} + O\!\left(r_{12}^{-3}\right)\;,\label{seq:rho_lin_approx}\\
\mathbf{m}^{(n)}(\mathbf{r}) &= \frac{\tilde{\rho}_b}{\rho_b}\mathbf{m}_2^{(n)} + O\!\left(u,r_{12}^{-3}\right) = \mathbf{m}_2^{(n)} + O\!\left(r_{12}^{-1}\right) \text{ for integers $n\ge 1$.} \label{seq:m_lin_approx}
\end{align}
Using these results and the single-body properties \eqref{seq:dipole_moment} and \eqref{seq:quadrupole_moment} in Eqs.~\eqref{seq:dipole_moment_2} and \eqref{seq:quadrupole_moment_2} for $j = 2$, we obtain more explicit formulas for the modified dipole and quadrupole moments
\begin{align}
\tilde{\mathbf{p}}_2 = \frac{\tilde{\rho}_b}{\rho_b}\mathbf{p}_2 + O\!\left(r_{12}^{-2}\right)
= \mathbf{p}_2 + \frac{\beta_\text{eff}}{2\pi\rho_b}\frac{\mathbf{r}_{12}\cdot\mathbf{p}_1}{r_{12}^2}\,\mathbf{p}_2 + O\!\left(r_{12}^{-2}\right) \;,\quad
\tilde{\mathbb{Q}}_2 = \mathbb{Q}_2 + O\!\left(r_{12}^{-1}\right)\;,
\end{align}
respectively. Exchanging the body indices $1$ and $2$, we also find the corresponding formulas for body~$1$
\begin{align}
\tilde{\mathbf{p}}_1 = \mathbf{p}_1 - \frac{\beta_\text{eff}}{2\pi\rho_b}\frac{\mathbf{r}_{12}\cdot\mathbf{p}_2}{r_{12}^2}\, \mathbf{p}_1 + O\!\left(r_{12}^{-2}\right) \;,\quad
\tilde{\mathbb{Q}}_1 = \mathbb{Q}_1 + O\!\left(r_{12}^{-1}\right)\;.
\end{align}
These, when used in Eq.~\eqref{seq:mod_rho_b_J_b}, imply
\begin{align} \label{seq:rho_b_J_b_final}
\tilde{\rho}_b &= \rho_b + \frac{\beta_\text{eff}}{2\pi}\left[\frac{\mathbf{r}_{12}\cdot\mathbf{p}_1}{r_{12}^2} - \frac{\beta_\text{eff}}{2\pi\rho_b}\frac{(\mathbf{r}_{12}\cdot\mathbf{p}_1)(\mathbf{r}_{12}\cdot\mathbf{p}_2)}{r_{12}^4} + \frac{\mathbf{r}_{12}\cdot\mathbb{Q}_1\mathbf{r}_{12}}{2r_{12}^4}\right] + O\!\left(r_{12}^{-3}\right) \nonumber\\
\tilde{\mathbf{J}}_b &= -\frac{\mu}{2\pi}\left[\frac{\mathbf{p}_1}{r_{12}^2} - \frac{2(\mathbf{r}_{12}\cdot\mathbf{p}_1)\mathbf{r}_{12}}{r_{12}^4}\right] + O\!\left(r_{12}^{-3}\right) = \mathbf{J}_1(\mathbf{r}_{12}) + O\!\left(r_{12}^{-3}\right)\;,
\end{align}
where $\mathbf{J}_1(\mathbf{r}_{12})$ denotes the current field generated by body~$1$ alone in the vicinity of body~$2$, as expressed in Eq.~\eqref{seq:current_exp}.

Finally, using Eqs.~\eqref{seq:rho_lin_approx} and \eqref{seq:rho_b_J_b_final}, the additional force on body~$2$ due to the presence of body~$1$ is obtained as
\begin{align}
\mathbf{F}_{12} &\equiv \mathbf{p}_2 + \int d^2\mathbf{r}\,\rho(\mathbf{r})\bm\nabla V_2 \nonumber\\
&= \left(1-\frac{\tilde{\rho}_b}{\rho_b}\right)\mathbf{p}_2 + \frac{1}{\rho_b}\sum_{a}\left(\tilde{\mathbf{J}}_b\cdot\mathbf{e}_a\right)\int d^2\mathbf{r}\,\left.\left[\left(\frac{\partial\rho_2^{(\mathbf{u})}}{\partial u_a}\right)\bm\nabla V_2\right]\right|_{\mathbf{u}=\mathbf{0}} + O\!\left(r_{12}^{-3}\right)\nonumber\\
&= - \frac{\beta_\text{eff}}{2\pi}\left[\frac{\mathbf{r}_{12}\cdot\mathbf{p}_1}{r_{12}^2} - \frac{\beta_\text{eff}}{2\pi\rho_b}\frac{(\mathbf{r}_{12}\cdot\mathbf{p}_1)(\mathbf{r}_{12}\cdot\mathbf{p}_2)}{r_{12}^4} + \frac{\mathbf{r}_{12}\cdot\mathbb{Q}_1\mathbf{r}_{12}}{2r_{12}^4}\right]\mathbf{p}_2 \nonumber\\
&\quad + \frac{1}{\rho_b}\mathbb{R}_2 \mathbf{J}_1(\mathbf{r}_{12}) + O\!\left(r_{12}^{-3}\right)\;,
\end{align}
where $\mathbb{R}_2$ is a linear-reponse tensor defined as
\begin{align}
\left(\mathbb{R}_2\right)_{ab} \equiv \int d^2\mathbf{r} \left.\left[\left(\frac{\partial\rho_2^{(\mathbf{u})}}{\partial u_b}\right)\left(\partial_a V_2\right)\right]\right|_{\mathbf{u}=\mathbf{0}}\;.
\end{align}
Alternatively, denoting by
\begin{align}
\mathbf{F}_2^{(\mathbf{u})} \equiv \int d^2\mathbf{r}\, \rho_2^{(\mathbf{u})}(\mathbf{r}) \bm\nabla V_2	
\end{align}
the steady-state force on body~$2$ in an active fluid of average density $\rho_b$ with a global diffusive current $\rho_b\mathbf{u}$, we can also write
\begin{align}
\left(\mathbb{R}_2\right)_{ab} = \left.\frac{\partial}{\partial u_b}\left[\mathbf{F}_2^{(\mathbf{u})}\cdot\mathbf{e}_a\right]\right|_{\mathbf{u}=\mathbf{0}}\;,
\end{align}
which reproduces Eq.~\eqref{eq:R2} of the main text.

It is natural to interpret $\mathbf{F}_{12}$ as the long-range force applied by body~$1$ on body~$2$. Moreover, $\mathbf{F}_{12}$ can be decomposed into two components
\begin{align}
\mathbf{F}_{12}^a &= - \frac{\beta_\text{eff}}{2\pi}\left[\frac{\mathbf{r}_{12}\cdot\mathbf{p}_1}{r_{12}^2} - \frac{\beta_\text{eff}}{2\pi\rho_b}\frac{(\mathbf{r}_{12}\cdot\mathbf{p}_1)(\mathbf{r}_{12}\cdot\mathbf{p}_2)}{r_{12}^4} + \frac{\mathbf{r}_{12}\cdot\mathbb{Q}_1\mathbf{r}_{12}}{2r_{12}^4}\right]\mathbf{p}_2 + O\!\left(r_{12}^{-3}\right)\;,\\
\mathbf{F}_{12}^s &= \frac{1}{\rho_b}\mathbb{R}_2 \mathbf{J}_1(\mathbf{r}_{12}) + O\!\left(r_{12}^{-3}\right)\;,
\end{align}
so that $\mathbf{F}_{12}^a$ acts only on body~$2$ with an asymmetric potential (which implies $\mathbf{p}_2 \neq 0$), while $\mathbf{F}_{12}^s$ is present even for fully symmetric bodies (for which $\mathbf{p}_2 = 0$). These reproduce the results shown in Eqs.~\eqref{eq:F12_nsym} and \eqref{eq:F12_sym} of the main text.

The torque applied by body~$1$ on body~$2$, denoted by $\bm\tau_{12}$, can be obtained in a similar manner. Let us denote by
\begin{align} \label{seq:tau_2_u}
\bm\tau_2^{(\mathbf{u})} \equiv \int d^2\mathbf{r}\, \rho_2^{(\mathbf{u})}(\mathbf{r})\,\mathbf{r}\times\bm\nabla V_2
\end{align}
the torque experienced by isolated body~$2$ in an active fluid of average density $\rho_b$ with a global diffusive current $\rho_b\mathbf{u}$. The self-torque of body~$2$ in the absence of the current is given by $\bm\tau_2 = \bm\tau_2^{(\mathbf{0})}$. Then we can write
\begin{align} \label{seq:tau_12_def}
\bm\tau_{12} &\equiv -\bm\tau_2 + \int d^2\mathbf{r} \,\rho(\mathbf{r})\,\mathbf{r}\times\bm\nabla V_2 \nonumber\\
&= \left(\frac{\tilde{\rho}_b}{\rho_b}-1\right)\bm\tau_2 + \frac{1}{\rho_b}\sum_a \left(\tilde{\mathbf{J}}_b\cdot\mathbf{e}_a\right)\left.\left(\frac{\partial\bm\tau_2^{(\mathbf{u})}}{\partial u_a}\right)\right|_{\mathbf{u}=\mathbf{0}} + O\!\left(r_{12}^{-3}\right)\;,
\end{align}
where Eqs.~\eqref{seq:rho_lin_approx} and \eqref{seq:tau_2_u} have been used to obtain the second equality. The expression can be further simplified by noting that
\begin{align} \label{seq:J_tau_2_coupling}
\sum_a \left(\tilde{\mathbf{J}}_b\cdot\mathbf{e}_a\right)\left.\left(\frac{\partial\bm\tau_2^{(\mathbf{u})}}{\partial u_a}\right)\right|_{\mathbf{u}=\mathbf{0}}
&= \left.\left[\left(\frac{\partial}{\partial u_x}\bm\tau_2^{(\mathbf{u})}\cdot\mathbf{e}_z\right)\left(\tilde{\mathbf{J}}_b\cdot\mathbf{e}_x\right) + \left(\frac{\partial}{\partial u_y}\bm\tau_2^{(\mathbf{u})}\cdot\mathbf{e}_y\right)\left(\tilde{\mathbf{J}}_b\cdot\mathbf{e}_y\right)\right]\right|_{\mathbf{u}=\mathbf{0}} \nonumber\\
&= \bm\gamma_2 \times \tilde{\mathbf{J}}_b\;,
\end{align}
where the vector $\bm\gamma_2$ is defined by
\begin{align}
\bm\gamma_2 \equiv \left.\begin{bmatrix}
\frac{\partial}{\partial u_y}\bm\tau_2^{(\mathbf{u})}\cdot\mathbf{e}_z\\
-\frac{\partial}{\partial u_x}\bm\tau_2^{(\mathbf{u})}\cdot\mathbf{e}_z	
 \end{bmatrix}\right|_{\mathbf{u}=\mathbf{0}}
 = \left.\left[\bm\nabla_\mathbf{u}\times\bm\tau_2^{(\mathbf{u})}\right]\right|_{\mathbf{u}=\mathbf{0}}\;,
\end{align}
which is also shown in Eq.~\eqref{eq:gamma} of the main text. Using Eqs.~\eqref{seq:rho_b_J_b_final} and \eqref{seq:J_tau_2_coupling} in Eq.~\eqref{seq:tau_12_def}, we obtain
\begin{align}
\bm\tau_{12} &= \frac{\beta_\text{eff}}{2\pi}\left[\frac{\mathbf{r}_{12}\cdot\mathbf{p}_1}{r_{12}^2} - \frac{\beta_\text{eff}}{2\pi\rho_b}\frac{(\mathbf{r}_{12}\cdot\mathbf{p}_1)(\mathbf{r}_{12}\cdot\mathbf{p}_2)}{r_{12}^4} + \frac{\mathbf{r}_{12}\cdot\mathbb{Q}_1\mathbf{r}_{12}}{2r_{12}^4}\right]\bm\tau_2 \nonumber\\
&\quad + \frac{1}{\rho_b}\bm\gamma_2\times \mathbf{J}_1(\mathbf{r}_{12}) + O\!\left(r_{12}^{-3}\right)\;.
\end{align}
This can be decomposed into two components
\begin{align}
\bm\tau_{12}^a &= \frac{\beta_\text{eff}}{2\pi}\left[\frac{\mathbf{r}_{12}\cdot\mathbf{p}_1}{r_{12}^2} - \frac{\beta_\text{eff}}{2\pi\rho_b}\frac{(\mathbf{r}_{12}\cdot\mathbf{p}_1)(\mathbf{r}_{12}\cdot\mathbf{p}_2)}{r_{12}^4} + \frac{\mathbf{r}_{12}\cdot\mathbb{Q}_1\mathbf{r}_{12}}{2r_{12}^4}\right]\bm\tau_2 + O\!\left(r_{12}^{-3}\right)\;,\\
\bm\tau_{12}^s &= \frac{1}{\rho_b}\bm\gamma_2 \times \mathbf{J}_1(\mathbf{r}_{12}) + O\!\left(r_{12}^{-3}\right)\;,
\end{align}
so that $\bm\tau_{12}^a$ acts only on body~$2$ with nonzero self-torque ($\bm\tau_2 \neq 0$), while $\bm\tau_{12}^s$ is present even when $\bm\tau_2 = 0$. These are in agreement with Eqs.~\eqref{eq:tau12_nsym} and \eqref{eq:tau12_sym} of the main text.

\begin{figure}
\includegraphics[width=0.7\textwidth]{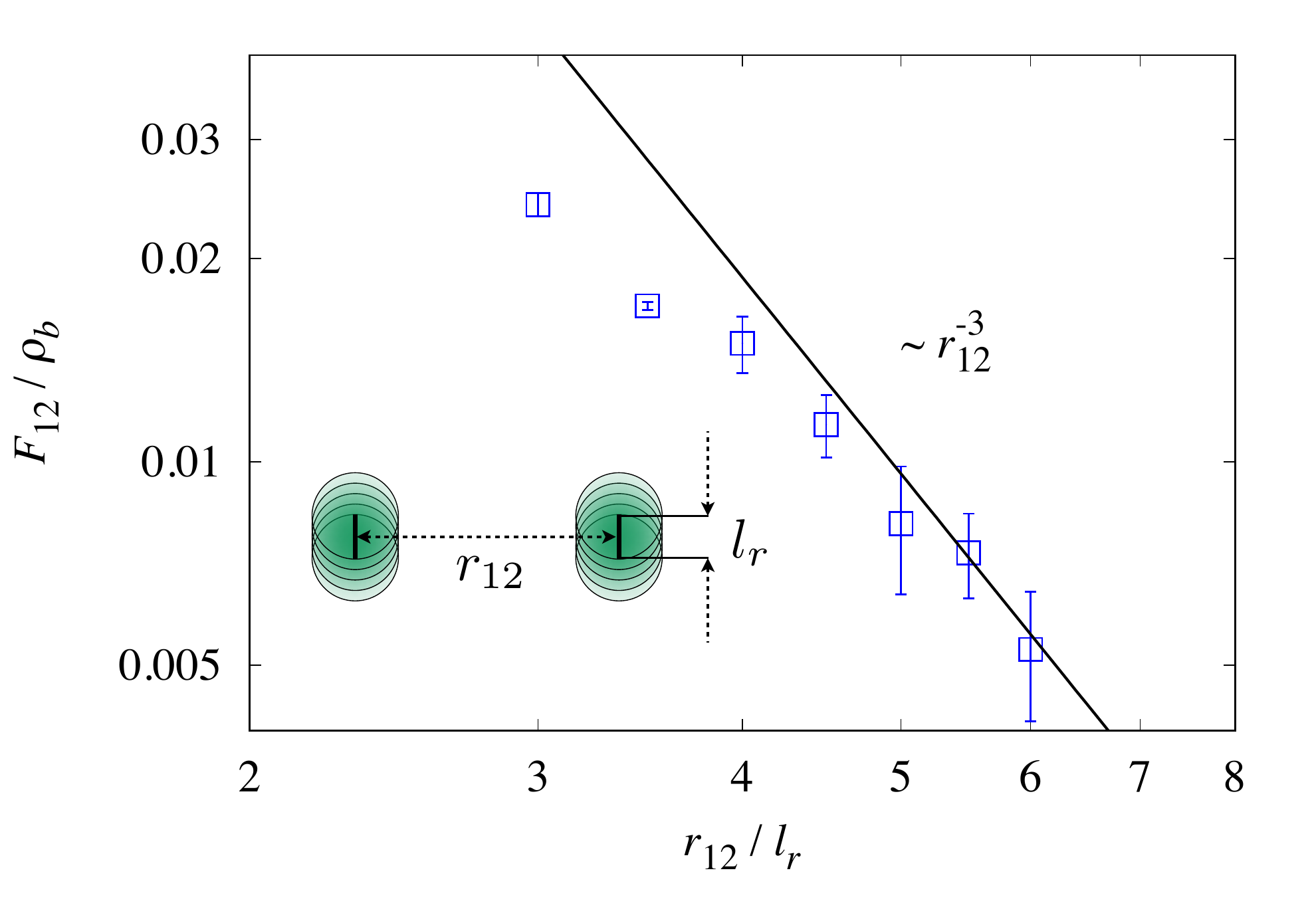}
\caption{\label{fig:figS2} The magnitude of the force $F_{12} \equiv |\mathbf{F}_{12}|$, normalized by the bulk density $\rho_b$ of the ambient active fluid, between a pair of rod-like bodies, which are parallel and separated by a normalized distance $r_{12}/l_r$, where $l_r$ is the run length of active particles. Each body consists of $5$ identical beads of radius $R_b = l_r$ and stiffness $\lambda = 2$, whose centers are aligned at equal distances from each other along a straight segment of length $l_r$. The active fluid is implemented by run-and-tumble particles of bulk density $\rho_b = 4800$ in a system of size $30 \times 30$ (see Sec.~\ref{app:parameters} for the other parameters and the units used). The simulation results (symbols) are consistent with a scaling behavior $F_{12} \sim r_{12}^{-3}$.}
\end{figure}

Repeating similar procedures, we can obtain interactions which are higher-order in $r_{12}^{-1}$. These are important for bodies whose leading-order interactions stem from higher-order multipoles. For instance, when bodies~$1$ and $2$ are both rod-like, all interactions at the orders discussed above are zero. The leading-order interaction between the two rods is expected to be controlled by quadrupole moments, which appear at the next order in the perturbative treatment, decaying as $r_{12}^{-3}$. This is numerically checked for the force in Fig.~\ref{fig:figS2} for a pair of rod-like bodies immersed in an active fluid of run-and-tumble particles. While numerical limitations prohibit us from verifying the result unambiguously, we find the numerics consistent with our prediction. 

\section{Adiabatic limit}
\label{app:adiabatic}

Here we elaborate on the adiabatic limit assumed for analyzing the dynamics of moving bodies. Consider first the case of an isolated body moving in an active fluid. As shown above, the active particles outside the body (where $V_j = 0$) behave as a diffusive fluid characterized by an effective diffusivity $D_\text{eff}$. Therefore, a density modulation, caused by the moving body, will spread the distance of the body~$l_j$ on a time scale $t_s$ set by $l_j \sim \sqrt{D_\text{eff} t_s}$. For the adiabatic approximation to hold, this process has to be fast enough so that the density near the body will be well approximated by that obtained from a non-moving body. Namely, we require that on this time scale the body should move a distance much smaller than its size, or $v_j t_s \ll l_j$. Here $v_j$ is the velocity of the body. This gives the condition for the validity of the approximation for a single moving body as
\begin{equation}
	v_j \ll \frac{D_\text{eff}}{l_j} \;.
\end{equation}
It is straightforward to extend these considerations for two interacting bodies $i$ and $j$. One finds that the condition is then given by
\begin{equation}
	\max \{v_i,v_j\} \ll \frac{D_\text{eff}}{r_{ij}} \;,
\end{equation}
where $r_{ij}$ is the distance between the bodies. We note that, despite these rather strict conditions, our results give much predictive power when used in numerics which do not obey these criteria.

\section{Perturbative solutions for multipole moments}
\label{app:perturbation}

In this section we present perturbative solutions in the strength of the potential for the dipole and quadrupole moments of body~$j$, assuming that changes in density occur on scales larger than the run length $l_r$. For simplicity, we focus on the case when the translational degrees of freedom are unaffected by the Gaussian white noise ($D_t = 0$). The generalization to the cases with $D_t \neq 0$ is straightforward but tedious.

To this end, we revisit Eq.~\eqref{seq:poisson} for the steady-state density profile $\rho_j$. Using Eq.~\eqref{seq:m_hier} iteratively, $\mathbf{m}_j^{(1)}$ and $\mathbf{m}_j^{(2)}$ can be expressed in terms of $\rho_j$ and $\mathbf{m}_j^{(3)}$. Then we divide both sides of Eq.~\eqref{seq:poisson} by $D_\text{eff}$ to obtain
\begin{align} \label{seq:poisson_perturb}
\nabla^2\rho_j &= - \beta_\text{eff}\bm\nabla\cdot(\rho_j\bm\nabla V_j) - \frac{\beta_\text{eff}l_r^2}{2}\sum_{a,\,b}\partial_a\partial_b(\partial_a V_j)(\partial_b\rho_j)\nonumber\\
&\quad + \frac{\beta_\text{eff}l_r^2}{4}\frac{\alpha+D_r}{\alpha+4D_r}\nabla^2\bm\nabla\cdot(\rho_j\bm\nabla V_j) + O(l_r^3\partial^3)\;. 
\end{align}
Here $O(l_r^3\partial^3)$ indicates a third or higher-order spatial derivative, with each derivative multiplied by $l_r$. Terms of this order can be identified by noting that, according to Eq.~\eqref{seq:m_hier}, reducing $\mathbf{m}_j^{(n)}$ to $\mathbf{m}_j^{(n-1)}$ always involves a factor of $l_r$, and that the operator $\mathbb{M}_j^{(1)}$ defined in Eq.~\eqref{seq:M_def} can be rewritten as
\begin{align}
\mathbb{M}_j^{(1)} = \frac{\mu}{\alpha+D_r}\bm\nabla\cdot\bm\nabla V_j
= \frac{\beta_\text{eff}l_r^2}{2}\bm\nabla\cdot\bm\nabla V_j\;.
\end{align}
When $l_r$ is much shorter than the length scale over which the density distribution changes (e.g. the penetration depth $d_j$), each $l_r\partial$ yields a small factor $l_r/d_j$. Thus $O(l_r^3\partial^3)$ is negligible compared to terms of the order $l_r^2\partial^2$. With this in mind, under the assumption of the weak potential ($\beta_\text{eff}V_j \ll 1$), we can solve Eq.~\eqref{seq:poisson_perturb} by a perturbation series $\rho_j = \sum_{k=0}^\infty \rho_{j,k}$ with $\rho_{j,k} \sim (\beta_\text{eff}V)^k$.

At zeroth order, the system is unaffected by the presence of body~$j$. Therefore $\rho_{j,0}$ is simply given by the bulk density $\rho_b$, which is constant over the entire space.

The next order can be obtained iteratively by Eq.~\eqref{seq:poisson_perturb}, which gives
\begin{align}
\nabla^2\rho_{j,1} &= -\rho_b\beta_\text{eff}\nabla^2 V_j + \frac{\rho_b\beta_\text{eff}l_r^2}{4}\frac{\alpha+D_r}{\alpha+4D_r}(\nabla^2)^2 V_j + O(l_r^3\partial^3)\;,
\end{align}
whose solution is
\begin{align} \label{seq:rho1}
\rho_{j,1}(\mathbf{r}) &= -\rho_b\beta_\text{eff} V_j + \frac{\rho_b\beta_\text{eff} \, l_r^2}{4}\frac{\alpha+D_r}{\alpha+4D_r}\nabla^2 V_j + O\!\left(\frac{l_r^3}{d_j^3}\right)\;.
\end{align}
Using this relation in Eq.~\eqref{seq:dipole_moment} yields a zero dipole moment, implying $\mathbf{p}_j = O(\beta_\text{eff}^2 V_j^2,\,l_r^3/d_j^3)$. On the other hand, combining Eq.~\eqref{seq:rho1} with Eq.~\eqref{seq:quadrupole_moment} and using the relations
\begin{align}
\int d^2\mathbf{r}'\,(\nabla'^2V_j)r'_a\partial'_b V_j &= -\int d^2\mathbf{r}'\,\left[(\partial'_a V_j)(\partial'_b V_j) - \frac{\delta_{ab}}{2}(\bm\nabla' V_j)^2\right]\;,\nonumber\\
\int d^2\mathbf{r}'\,V_j r'_a \partial'_b V_j &= -\frac{\delta_{ab}}{2}\int d^2\mathbf{r}'\,V_j^2\:,
\end{align}
one finds the quadrupole moment
\begin{align} \label{seq:quadrupole2}
(\mathbb{Q}_j)_{ab} = -\rho_b \beta_\text{eff} l_r^2\, \frac{\alpha+7D_r}{\alpha+4D_r}\int d^2\mathbf{r}\,\left[(\partial_a V_j)(\partial_b V_j)-\frac{\delta_{ab}}{2}(\bm\nabla V_j)^2\right] + O\!\left(\beta_\text{eff}^2V_j^2,\,\frac{l_r^3}{d_j^3}\right)\;.
\end{align}
Thus, if $|\bm\nabla V_j|_\text{max}$ denotes the maximal force applied by the potential of body~$j$, the quadrupole moment satisfies
\begin{align}
	|\mathbb{Q}_j| \sim l_r^2\,|\bm\nabla V_j|_\text{max}^2
\end{align}
in the limit of weak potential and small run length.

\begin{figure}
\includegraphics[width=0.7\textwidth]{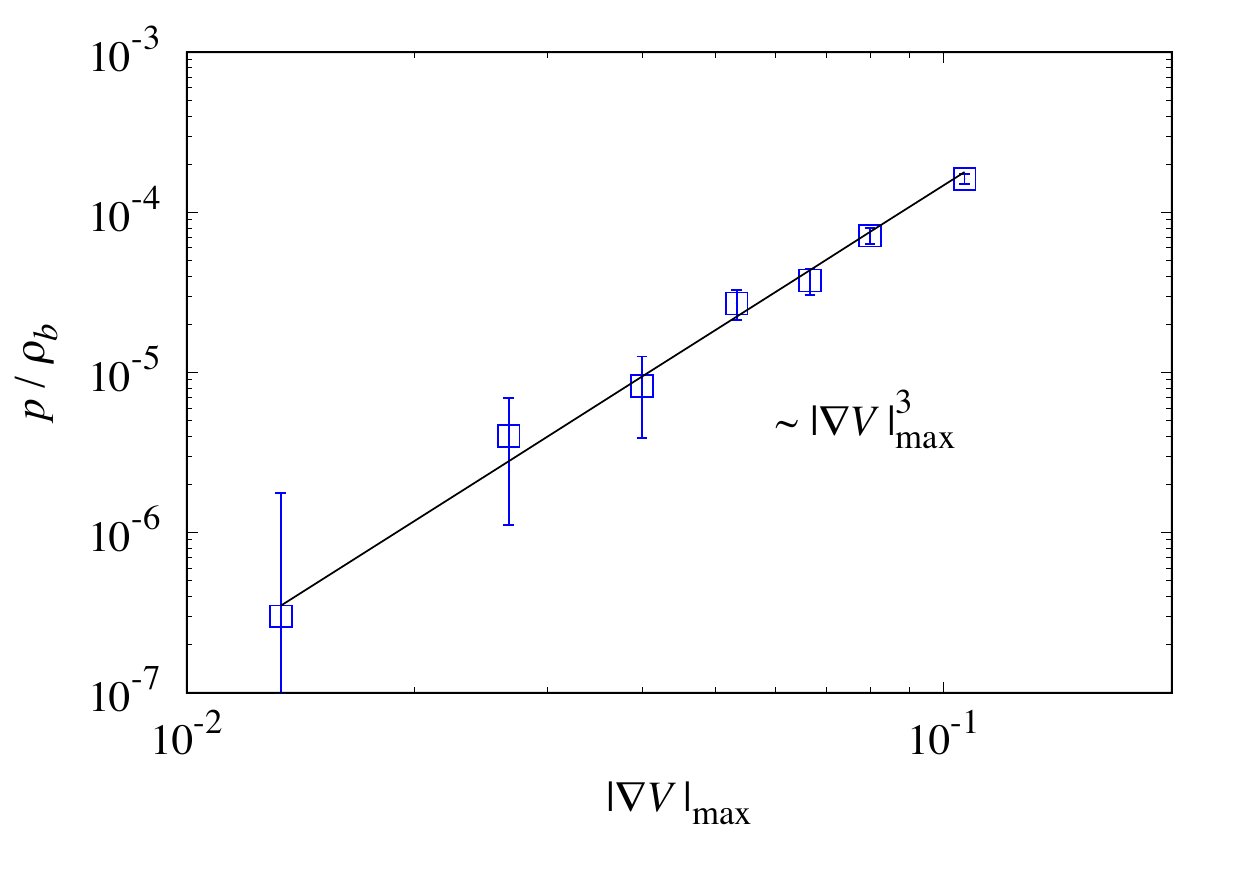}
\caption{\label{fig:figS3} The dipole moment $p$, normalized by the bulk density $\rho_b$ of the ambient active fluid, of an asymmetric body as a function of the maximal force $|\bm\nabla V|_\text{max}$ applied by its potential. The potential is a skewed Schwartz bell, which is $V(r,\theta) = \lambda \, e^{-1/\left(1-r^{2}\right)}\left(1+r\cos\theta\right)$ for $r \le 1$ and zero for $r \ge 1$. The dipole moment lies along the $x$-axis, the only asymmetric direction of the potential. The active fluid consists of noninteracting run-and-tumble particles with $v = \mu = 1$, $\alpha=0.3$, and $\rho_b = 3.75$ in a system of size $40 \times 40$ (in arbitrary units). The simulation results (symbols) are compared with the prediction (solid line) of Eq.~\eqref{seq:dipole_moment_perturb}.}
\end{figure}

To obtain the leading dipole moment, we proceed to the next order. From Eqs.~\eqref{seq:poisson_perturb} and \eqref{seq:rho1}, one obtains
\begin{align} \label{seq:poisson_rho_2}
\nabla^2\rho_{j,2} &= \frac{\rho_b \beta_\text{eff}^2}{2}\nabla^2 V_j^2 - \frac{\rho_b \beta_\text{eff}^2 l_r^2}{8}\frac{\alpha+D_r}{\alpha+4D_r}(\nabla^2)^2 V_j^2 - \frac{\rho_b \beta_\text{eff}^2 l_r^2}{4}\frac{\alpha+D_r}{\alpha+4D_r}\bm\nabla\cdot(\bm\nabla V_j)\nabla^2 V_j \nonumber\\
&\quad + \frac{\rho_b \beta_\text{eff}^2 l_r^2}{2}\sum_{a,\,b}\partial_a\partial_b(\partial_a V_j)(\partial_b V_j) + O(l_r^3\partial^3)\;.
\end{align}
Noting
\begin{align}
\int d^2\mathbf{r}' \, \ln \left|\mathbf{r}-\mathbf{r}'\right| &\bm\nabla'\cdot(\bm\nabla' V_j)\nabla'^2 V_j
= \int d^2\mathbf{r}' \, \frac{(\mathbf{r}-\mathbf{r}')\cdot\bm\nabla' V_j}{|\mathbf{r}-\mathbf{r}'|^2}\,\nabla'^2 V_j \nonumber\\
&= -\sum_{a,\,b} \int d^2\mathbf{r}' \, (\partial'_a V_j)\,\partial'_a \left[\frac{(r_b-r'_b) (\partial'_b V_j)}{|\mathbf{r}-\mathbf{r}'|^2}\,\nabla'^2 V_j\right] \nonumber\\
&= \int d^2\mathbf{r}' \, \left\{\frac{(\bm\nabla' V_j)^2}{|\mathbf{r}-\mathbf{r}'|^2} - \frac{2[(\mathbf{r}-\mathbf{r}') \cdot (\bm\nabla' V_j)]^2}{|\mathbf{r}-\mathbf{r}'|^4} -\frac{\mathbf{r}-\mathbf{r}'}{2|\mathbf{r}-\mathbf{r}'|^2}\cdot\bm\nabla'(\bm\nabla'V_j)^2\right\} \nonumber\\
&= \int d^2\mathbf{r}' \, \left\{\frac{3(\bm\nabla' V_j)^2}{2|\mathbf{r}-\mathbf{r}'|^2} - \frac{2[(\mathbf{r}-\mathbf{r}') \cdot (\bm\nabla' V_j)]^2}{|\mathbf{r}-\mathbf{r}'|^4}\right\}\;,\\
\int d^2\mathbf{r}' \, \ln \left|\mathbf{r}-\mathbf{r}'\right| &\partial'_a\partial'_b(\partial'_a V_j)(\partial'_b V_j) = \int d^2\mathbf{r}' \, \left\{\frac{(\bm\nabla' V_j)^2}{|\mathbf{r}-\mathbf{r}'|^2} - \frac{2[(\mathbf{r}-\mathbf{r}') \cdot (\bm\nabla' V_j)]^2}{|\mathbf{r}-\mathbf{r}'|^4}\right\}\;,
\end{align}
the solution of Eq.~\eqref{seq:poisson_rho_2} is found to be
\begin{align} \label{seq:rho2}
\rho_{j,2}(\mathbf{r}) &= \frac{\rho_b\beta_\text{eff}^2}{2}\,V_j^2 - \frac{\rho_b \beta_\text{eff}^2 l_r^2}{8}\frac{\alpha+D_r}{\alpha+4D_r}\nabla^2 V_j^2 +\frac{\rho_b \beta_\text{eff}^2 l_r^2}{16\pi} \frac{\alpha+13D_r}{\alpha+4D_r} \int d^2\mathbf{r}' \,\frac{(\bm\nabla'V_j)^2}{|\mathbf{r}-\mathbf{r}'|^2}\nonumber\\
&\quad - \frac{\rho_b \beta_\text{eff}^2 l_r^2}{4\pi} \frac{\alpha+7D_r}{\alpha+4D_r}\int d^2\mathbf{r}'\,\frac{[(\mathbf{r}-\mathbf{r}')\cdot(\bm\nabla' V_j)]^2}{|\mathbf{r}-\mathbf{r}'|^4} + O\!\left(\frac{l_r^3}{d_j^3}\right)\;.
\end{align}
Using Eqs.~\eqref{seq:dipole_moment}, \eqref{seq:rho2}, and
\begin{align}
\int d^2\mathbf{r}\, (\nabla^2 V_j^2)\bm\nabla V_j &= \int d^2\mathbf{r}\, (\bm\nabla \cdot 2 V_j \bm\nabla V_j)\bm\nabla V_j = -2\sum_a \int d^2\mathbf{r}\, V_j (\partial_a V_j)(\bm\nabla \partial_a V_j) \nonumber\\
&= -\int d^2\mathbf{r}\, V_j \bm\nabla (\bm\nabla V_j)^2 = \int d^2\mathbf{r}\, (\bm\nabla V_j)^2 \bm\nabla V_j\;,
\end{align}
the dipole moment is obtained as
\begin{align} \label{seq:dipole_moment_perturb}
\mathbf{p}_j &= \frac{\rho_b \beta_\text{eff}^2 l_r^2}{8} \frac{\alpha+D_r}{\alpha+4D_r} \int d^2\mathbf{r}\,(\bm\nabla V_j)^2\bm\nabla V_j -\frac{\rho_b \beta_\text{eff}^2 l_r^2}{16\pi} \frac{\alpha+13D_r}{\alpha+4D_r} \int d^2\mathbf{r}\, (\bm\nabla V_j) \int d^2\mathbf{r}'\,\frac{(\bm\nabla'V_j)^2}{|\mathbf{r}-\mathbf{r}'|^2}\nonumber\\
&\quad +\frac{\rho_b \beta_\text{eff}^2 l_r^2}{4\pi} \frac{\alpha+7D_r}{\alpha+4D_r} \int d^2\mathbf{r}\, (\bm\nabla V_j) \int d^2\mathbf{r}'\,\frac{[(\mathbf{r}-\mathbf{r}')\cdot(\bm\nabla' V_j)]^2}{|\mathbf{r}-\mathbf{r}'|^4} + O\!\left(\beta_\text{eff}^3 V_j^3,\,\frac{l_r^3}{d_j^3}\right)\;.
\end{align}
Thus the magnitude of the dipole moment satisfies
\begin{align}
	p_j \sim l_r^2\,|\bm\nabla V_j|_\text{max}^3
\end{align}
in the limit of weak potential and small run length. A numerical verification of the scaling behavior $p_j \sim |\bm\nabla V_j|_\text{max}^3$ is shown in Fig.~\ref{fig:figS3}.

\end{spacing}

\bibliography{interaction_active_fluid}

\end{document}